\documentclass{aa}  
\usepackage{graphicx}
\usepackage[dvipsnames]{xcolor}
\usepackage[varg]{txfonts}
\usepackage[normalem]{ulem}
\usepackage{amsmath}

\bibpunct{(}{)}{;}{a}{}{,}
\graphicspath{{./}{plots/}}

\usepackage{xpatch}
\usepackage{hyperref}
\hypersetup{
	colorlinks=true,
	breaklinks=true,
	citecolor=blue,
	allcolors=blue,
	frenchlinks=true
}

\makeatletter
\xpatchcmd\NAT@citex
{%
	\@citea\NAT@hyper@{%
		\NAT@nmfmt{\NAT@nm}%
		\hyper@natlinkbreak{\NAT@aysep\NAT@spacechar}{\@citeb\@extra@b@citeb}%
		\NAT@date
	}%
}
{%
	\@citea
	\NAT@nmfmt{\NAT@nm}%
	\NAT@aysep\NAT@spacechar
	\NAT@hyper@{\NAT@date}%
}
{}{}
\xpatchcmd\NAT@citex
{%
	\@citea\NAT@hyper@{%
		\NAT@nmfmt{\NAT@nm}%
		\hyper@natlinkbreak{\NAT@spacechar\NAT@@open\if*#1*\else#1\NAT@spacechar\fi}%
		{\@citeb\@extra@b@citeb}%
		\NAT@date
	}%
}
{
	\@citea
	\NAT@nmfmt{\NAT@nm}%
	\NAT@spacechar\NAT@@open\if*#1*\else#1\NAT@spacechar\fi
	\NAT@hyper@{\NAT@date}%
}
{}{}
\makeatother

\newcommand{\bs}[1]{\boldsymbol{#1}}
\newcommand{\bcdot}{\boldsymbol{\cdot}}
\newcommand{\bnabla}{\boldsymbol{\nabla}}

\newcommand{\kB}{k_\mathrm{B}}
\newcommand{\mP}{m_\mathrm{p}}

\newcommand{\Xmag}{X_\mathrm{mag}}
\newcommand{\Xmaginit}{X_\mathrm{mag,0}}
\newcommand{\Xcr}{X_\mathrm{cr}}
\newcommand{\Xcrinit}{X_\mathrm{cr,0}}
\newcommand{\Xkin}{X_\mathrm{kin}}
\newcommand{\Xkininit}{X_\mathrm{kin,0}}
\newcommand{\kcr}{\kappa_\mathrm{cr}}
\newcommand{\keff}{\kappa_\mathrm{eff}}
\newcommand{\Pth}{P_\mathrm{th}}
\newcommand{\Pmag}{P_\mathrm{mag}}
\newcommand{\Pkin}{P_\mathrm{kin}}
\newcommand{\Pcr}{P_\mathrm{cr}}
\newcommand{\tcool}{t_\mathrm{cool}}
\newcommand{\tff}{t_\mathrm{ff}}
\newcommand{\tcr}{t_\mathrm{cr}}
\newcommand{\tcol}{t_\mathrm{collapse}}
\newcommand{\NH}{N_\mathrm{H}}
\newcommand{\nH}{n_\mathrm{H}}
\newcommand{\vA}{\varv_\mathrm{a}}
\newcommand{\vcr}{\varv_\mathrm{cr}}
\newcommand{\ecr}{\varepsilon_\mathrm{cr}}
\newcommand{\fcr}{f_\mathrm{cr}}
\newcommand{\gcr}{\gamma_\mathrm{cr}}
\newcommand{\unitKappa}{\mathrm{cm^{2}~s^{-1}}}
\newcommand{\unitRho}{\mathrm{g~cm^{-3}}}
\newcommand{\unitT}{\mathrm{K}}

\begin{document} 

\title{CRexit: How different cosmic ray transport modes affect thermal instability in the circumgalactic medium}

\author{M. Weber\inst{1,2}
      \and
      T. Thomas\inst{1}
      \and
      C. Pfrommer\inst{1}    
      \and
      R. Pakmor\inst{3}
      }

\institute{
Leibniz Institute for Astrophysics Potsdam (AIP), An der Sternwarte 16, D-14482 Potsdam, Germany\\
\email{maweber@aip.de}
\and
Institute for Physics and Astronomy, University Potsdam, Karl-Liebknecht-Str. 24/25, 14476 Potsdam, Germany
\and
Max-Planck Institute for Astrophysics (MPA), Karl-Schwarzschild-Str. 1, D-85748 Garching, Germany
}

\date{\today}

\abstract{ 
The circumgalactic medium (CGM) plays a critical role in galaxy evolution, influencing gas flows, feedback processes, and galactic dynamics. Observations have shown a substantial cold gas reservoir in the CGM, but the mechanisms driving its formation and evolution remain unclear. Cosmic rays (CRs), as a source of non-thermal pressure, are increasingly recognised as key regulators of cold gas dynamics.
This study explores how CRs affect cold clouds that condense from the hot CGM through thermal instability (TI). Using three-dimensional CR magnetohydrodynamics simulations with the moving-mesh code \textsc{Arepo}, we assessed the impact of various CR transport models on cold gas evolution.
Under purely advective CR transport, CR pressure significantly suppressed the collapse of thermally unstable regions, altering the CGM's structure. 
In contrast, our realistic CR transport models revealed that CRs escape collapsing regions via anisotropic streaming and diffusion along magnetic fields, reducing their ability to prevent collapse and diminishing their impact on the thermal structure of the cold CGM. 
The ratio of the CR escape timescale to the cloud collapse timescale emerged as a critical factor in determining the influence of CRs on TI. The CRs remained confined within cold clouds when effective CR diffusion was slow, thereby maximising their pressure support and inhibiting collapse. The fast and effective CR diffusion realised in our two-moment CR-magnetohydrodynamics model facilitated rapid CR escape, diminishing their stabilising effect. This realistic CR transport model shows a wide dynamic range of the effective CR diffusion coefficient; its CR-energy-weighted median ranges from $10^{29}$ to $10^{30}\,\unitKappa$ for thermally  to CR-dominated atmospheres, respectively.
In addition to these CR transport-related effects, we demonstrated that a high numerical resolution is crucial, as it is necessary to avoid spuriously large clouds formed in low-resolution simulations, which would result in overly long CR escape times and artificially amplified CR pressure support.}


\titlerunning{How different cosmic ray transport modes affect thermal instability}
\authorrunning{Weber et al.}
\maketitle

\section{Properties of the circumgalactic medium}
The CGM encompasses an extensive gas-rich envelope surrounding most galactic discs, spanning from dwarfs to massive ellipticals and extending to the virial radius and beyond \citep{Tumlinson2017}. Observational studies suggest that the CGM can harbour a gas mass surpassing that of a galactic disc, underscoring its significance as a dynamic reservoir. The CGM plays a pivotal role in galactic evolution, serving both as a source of gas accretion for star formation and as a sink for energy, mass, and momentum, which are injected by feedback-driven galactic winds \citep{Faucher-Giguere2023}. 

The CGM exhibits a complex multiphase structure characterised by significant variations in density, temperature, and ionisation state \citep{Chen2010, Prochaska2011, Sameer2024}. It consists of a hot diffuse phase with temperatures exceeding $10^6\,\unitT$, as observed in X-ray studies \citep{Anderson2010, Gupta2012}; a warm phase at around $10^5 \,\unitT$, traced by UV absorption lines \citep{Wakker2012}; and a cold phase near $10^4 \,\unitT$ \citep{Wisotzki2018}, which may account for up to 50\% of the galactic halo's baryonic gas mass \citep{Werk2014}. Cold dense gas is present in various environments such as multiphase outflows \citep{Veilleux2020}, high-velocity clouds \citep{Bregman1980, Wakker1997, Putman2003, Maller2004, Richter2017}, and accretion streams that supply material to galactic discs \citep{Rubin2010, Stern2024}.
Despite its significance, the origin of the cold phase within the CGM remains a fundamental question in the context of galactic evolution that is yet to be resolved.

A common idea regarding the origin of this phase is that the cold gas is transported from the interstellar medium (ISM) into the CGM via galactic winds and outflows triggered by supernovae (SNe) and active galactic nuclei (AGNe). In such turbulent settings, cold clouds can be disrupted and fragmented by the ambient hot wind \citep{McCourt2018, Gronke2018, Sparre2019, Sparre2020, Li2020}, reshaped or protected by magnetic draping layers \citep{Dursi2008, McCourt2015, Jung2023, Ramesh2024}, or dissociated by the intense ultraviolet (UV) radiation field emanating from the galactic centre \citep{Decataldo2019}, which are processes that significantly alter the survival characteristics of cold clouds. Another theory suggests that the gas condenses locally via TI \citep{Field1965, McCourt2012, Sharma2012}, a process initiated by local density perturbations, and eventually this condensation results in the precipitation of cold clouds onto the galactic disc \citep{Voit2015, Voit2017}. 

Observations indicate that the cold phase of the CGM is out of pressure equilibrium with the surrounding hot medium \citep{Werk2014}, which may suggest additional non-thermal pressure sources \citep{McQuinn2018}\footnote{These constraints on the non-thermal pressure are upper limits because the cool-phase CGM density was derived from integrated column densities of all ionic components \citep{Werk2014}. Because of density clumping and observational biases that favour intermediate-state ions (\ion{C}{iii}, \ion{Si}{iii}, etc.), the inferred mean density of the cold phase may be underestimated, as suggested by the density distribution of individual absorption systems \citep{Zahedy2019, Qu2023}.} to maintain the overall stability of the CGM \citep{Tumlinson2017}. Cosmic rays, as a pervasive and energetically significant component of galactic ecosystems, are increasingly recognised for their role in shaping the dynamics of the CGM \citep{Ferriere2001, Owen2023, Ruszkowski2023}. In the galactic disc, these relativistic particles can provide non-thermal pressure support comparable to or exceeding that of thermal gas, magnetic fields, or turbulence \citep{Boulares1990, Naab2017}, thus influencing large-scale dynamics and the stability of gas phases. The CRs acquire their high energies through diffusive acceleration processes at supernova-induced shocks \citep{Marcowith2016} or from AGN-driven feedback \citep{Guo2008, Jacob2017a, Jacob2017b} and are transported from their sources within the ISM into the CGM by galactic winds, as shown in global galaxy simulations \citep{Uhlig2012, Salem2014, Ruszkowski2017, Buck2020, Dashyan2020, Hopkins2020, Thomas2023, Thomas2024} or in local high-resolution ISM setups \citep{Girichidis2016, Farber2018, Sike2024}.

Although previous studies have highlighted the potential of CRs to modulate TI \citep{Butsky2020, Beckmann2022} and to influence the multiphase structure of the CGM \citep{Butsky2018, Ji2020, Thomas2024}, the interplay between CR transport mechanisms and the resulting cold gas morphology remains poorly understood. Cosmic rays have the ability to both suppress and enhance TI depending on the underlying transport physics and the properties of the ambient medium \citep{Wiener2017a, Hopkins2021}. Furthermore, CR heating driven by interactions with Alfv\'en waves may alter cooling rates, thereby affecting the conditions under which cold gas forms and survives \citep{Zweibel2013, Ruszkowski2017}.

Addressing the challenges related to CR modulated TI in the CGM necessitates detailed simulations that capture the full complexity of CR physics. By systematically exploring the effects of varying CR pressure, transport mechanisms, and transport speeds, we aim to develop a comprehensive understanding of how CRs shape the CGM. In this study, we focus on the impact of CRs on the morphology and observable properties of cold gas formed via TI in CGM environments. To achieve this, we performed a series of three-dimensional cosmic ray magnetohydrodynamics (CRMHD) simulations using the moving-mesh code \textsc{Arepo}. We examined the outcomes of simulations implementing different CR transport mechanisms across a range of environments, from those dominated by thermal pressure to those heavily influenced by CR pressure. The outline of this paper is as follows. In Sec.~\ref{sec:crmodulatedTI}, we provide a brief explanation of the theoretical background of the process of TI and the nature of CR physics. In Sec.~\ref{sec:numericalsetup}, we present our numerical and physical setup along with the initial conditions for the simulated CGM-like environments. We discuss the suppression of TI employing purely advective CR transport in Sec.~\ref{sec:advectivecrs}. In Sec.~\ref{sec:2momentcrs}, we investigate and analyse the revival of TI in the context of two-moment CR transport. In Sec.~\ref{sec:crtransportspeed}, we detail the importance of CR transport speed on the onset of TI and compare the purely diffusive CR transport to the two-moment approach in this context. Section~\ref{sec:resolution} is dedicated to emphasising the critical role of numerical resolution in simulating CRs in the CGM. We close this paper with a discussion of our results and the limitations of the employed setup in Sec.~\ref{sec:discussion}, and we present our conclusion in Sec.~\ref{sec:conclusions}. In Appendix~\ref{app:coolinglengthresolution}, we address the issue of numerical resolution in a simulation of TI, while we study the influence of the ratio of cooling to free-fall time on the emerging multiphase structure in our TI setup in Appendix~\ref{app:tiwithourcrs}.

\section{Cosmic ray modulated thermal instability}
\label{sec:crmodulatedTI}
In this section, we provide a concise summary of the key physical processes that regulate CR-modulated TI. We also outline the metrics used for our analysis throughout this study.

\begin{figure*}
    \centering
    \includegraphics[width=0.49\linewidth]{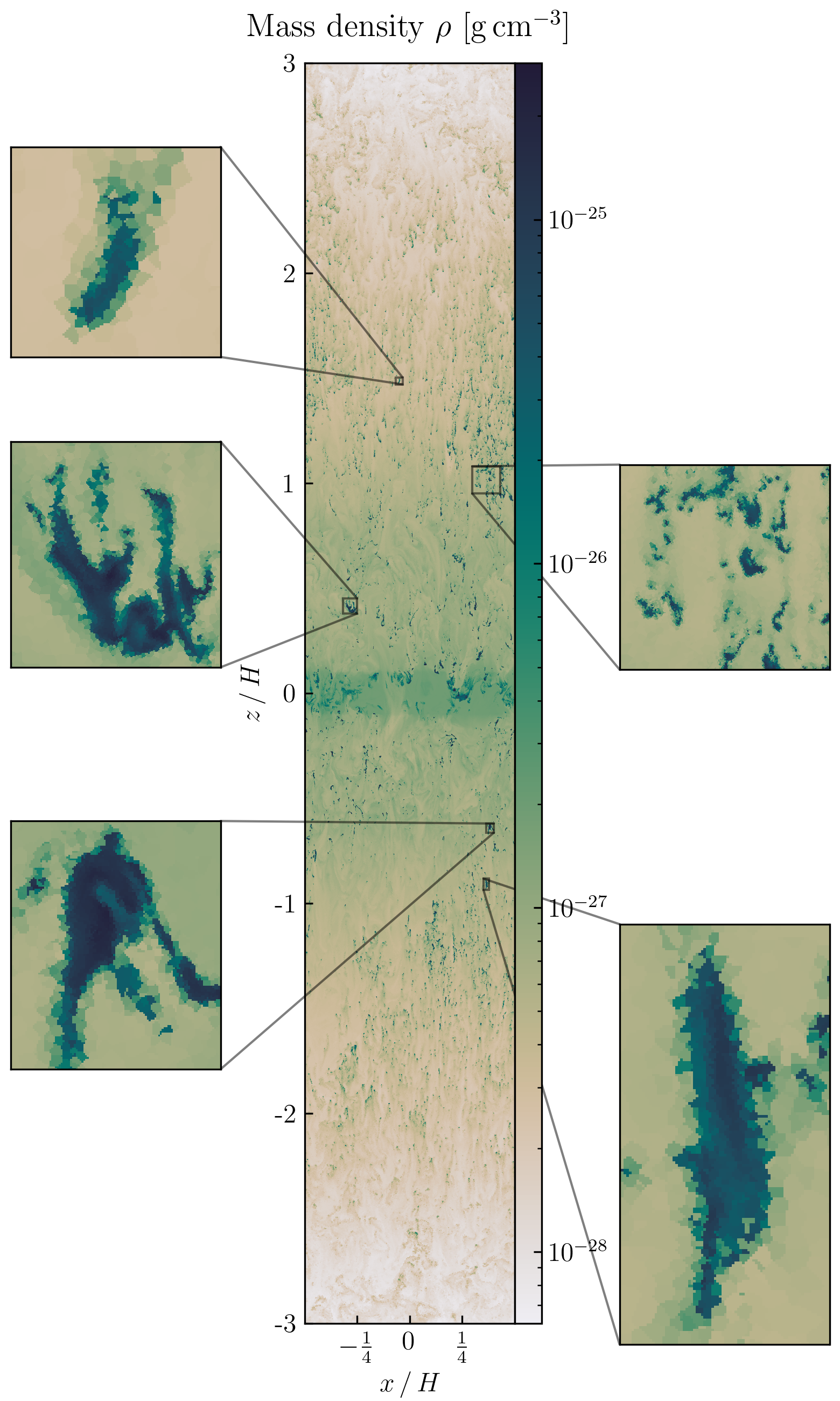}
    \includegraphics[width=0.49\linewidth]{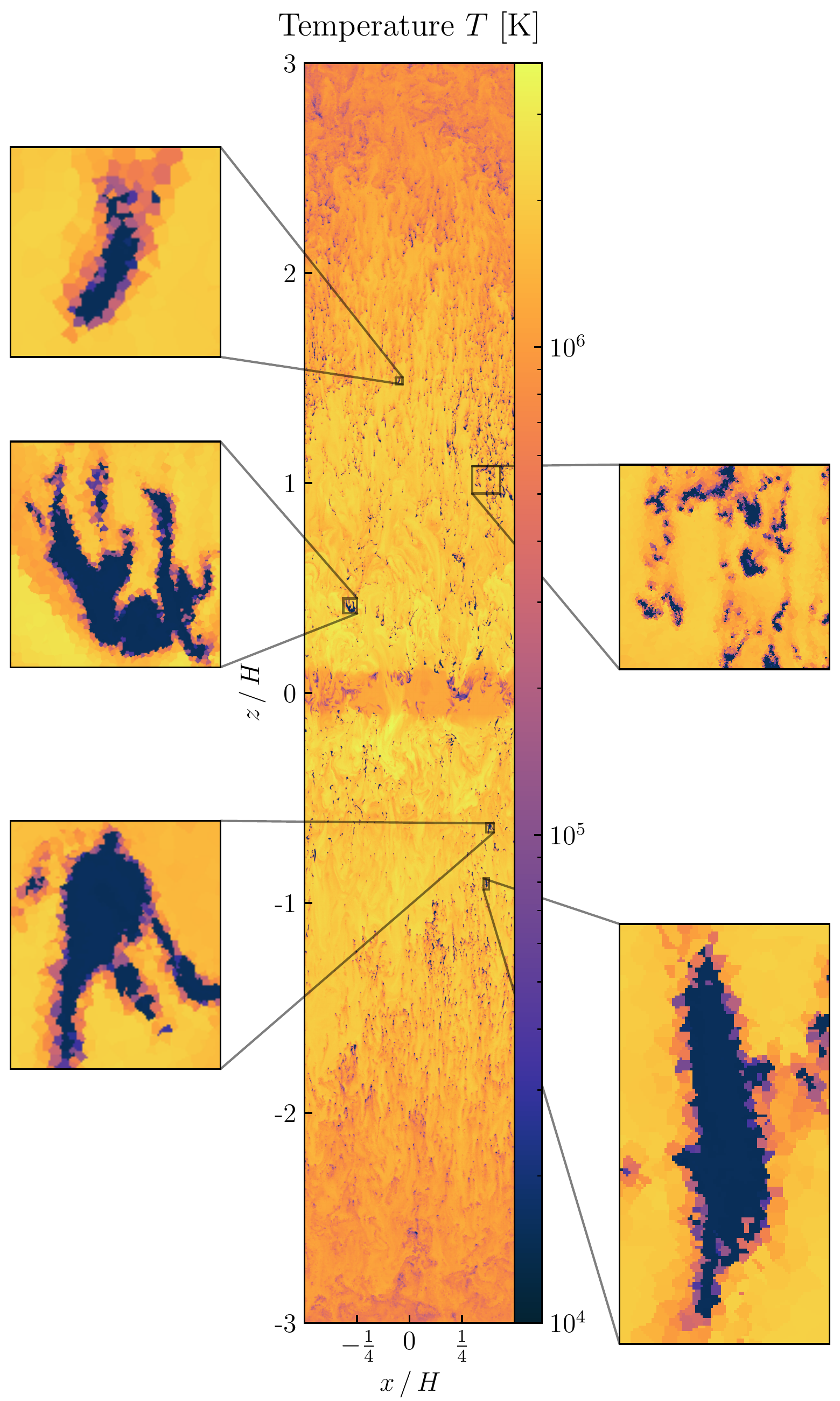}
    \caption{Simulation domain overview. The figure shows a slice through the $x$-$z$ plane at $y=0$ of our tall-box setup. The box extends from $-3H$ to $+3H$ in the $z$-direction and from $-0.5H$ to $+0.5H$ in the $x$- and $y$-directions, where $H=30.11$ kpc. Periodic boundary conditions are applied in the $x$- and $y$-directions, while the $z$-direction features an outflow boundary condition. The simulation is conducted without CRs and assumes a target mass of $m_\mathrm{target} = 22\,\mathrm{M}_\odot$. The computational domain is initialised with $256\times256\times1536$ equally spaced mesh-generating points. The left elongated panel illustrates the resulting mass density, and the right elongated panel displays the temperature. The five smaller panels provide magnified views of various structures of different sizes, shapes, densities, and temperatures formed due to TI.}
    \label{fig:setup}
\end{figure*}

\subsection{Thermal instability}
Thermal instability is a self-amplifying process driven by the dependence of the cooling function on temperature and density in astrophysical plasmas \citep{Field1965}. In an initially stable gas distribution, small density enhancements accelerate cooling, leading to a reduction in thermal pressure support relative to the surrounding hot environment. This pressure imbalance causes the over-dense regions to condense further, increasing the local density and amplifying the cooling process. This positive feedback loop results in the growth of density perturbations, culminating in the formation of cold, dense structures within the medium.

The cooling time-to-free-fall time ratio, $\tau=\tcool/\tff$, has been identified as a critical parameter for initiating this runaway process \citep{McCourt2012, Sharma2010b, Sharma2012, Voit2015}. In a stratified atmosphere, as cooling gas parcels sink deeper into the gravitational potential, the ambient thermal pressure increases, inducing compressive heating. For $\tau \lesssim 1$, cooling dominates over compressive heating, allowing TI to grow and form dense condensations. However, when $\tau \gtrsim 10$, the onset of TI is suppressed as a result of buoyancy.

Several additional factors influence the development of TI beyond $\tau$. For instance, the amplitude of initial density perturbations can significantly impact its growth \citep{Choudhury2019, Esmerian2021}. Magnetic fields, which modify gas dynamics, can either suppress or enhance instability depending on their configuration \citep{Ji2018, Fournier2024}. Similarly, turbulence \citep{Voit2018, Mohapatra2023}, metallicity \citep{Das2021}, thermal conduction \citep{Koyama2004, McCourt2012, Wagh2014}, and of course, CRs \citep{Ruszkowski2023} all play essential roles in regulating the onset and progression of TI.

Figure~\ref{fig:setup} presents 2D slices of gas mass density (left panel) and temperature (right panel) from a simulation without CR physics. Magnified insets show regions where TI is actively developing. These insets highlight the size and structure of the emerging cold gas clouds. The simulation employs $\tau = 0.3$, corresponding to a regime where radiative cooling dominates over compressive heating. This configuration serves as the fiducial case for this study. We confirm findings of previous studies, detailed in Appendix~\ref{app:tiwithourcrs}, which explore TI in the absence of CR physics. Building on this foundation, the focus of the present work shifts to the influence of CRs on the formation and evolution of TI in CGM-like environments.

\subsection{Cosmic ray physics}
We used the CRMHD framework developed by \citet{Thomas2019}. While we summarise the key physical concepts of this two-moment method for CR transport here, we refer the reader to \citet{Thomas2023} for a more detailed and illustrative description of the underlying CR physics.

The transport of CRs, being charged particles, is influenced by their interaction with magnetic fields. Their gyroradii ($\sim$0.25~AU for giga-electronvolt CRs in microgauss magnetic fields) are much smaller than the typical length scales of the CGM dynamics we are investigating here. Consequently, we can assume that they are tied to magnetic field lines. Because magnetic field lines also follow the gas motion, we can also readily assume that CRs are advected with the gas flow. On top of this, CRs are free to propagate anisotropically along magnetic field lines but interact with small perturbations of the magnetic field which determine how fast CRs are transported along magnetic field lines. Notably, CRs can excite Alfv\'en waves which are plasma waves that propagate at the Alfv\'en speed, $\vA$. When the mean CR transport speed exceeds the local Alfv\'en speed ($\vcr > \vA$), CRs will excite and amplify resonant Alfv\'en waves through the gyroresonant instability \citep{Kulsrud1969, Shalaby2023}. These Alfv\'en waves act as magnetic perturbations that affect the CR’s gyromotion along the field lines leading to effective scattering of the CRs and a transfer of energy and momentum to the waves. Conversely, when the mean CR transport speed is lower than the Alfv\'en speed ($\vcr < \vA$), the excitation of the gyroresonant instability is suppressed. Instead, CRs can gain energy from Alfv\'en waves through a second-order Fermi process \citep{Ko1992} resulting in their acceleration.

The scattering of CRs off Alfv\'en waves redistributes their momentum and energy, effectively limiting their transport speed to a value close to $\vA$. In regions with high scattering frequencies, $\nu_\mathrm{cr}$, CRs remain strongly coupled to the waves, resulting in highly constrained motion along the magnetic field lines. This regime is characterised by CR streaming, where CR propagation is regulated by the local magnetic field properties and the efficiency of wave generation and damping. In contrast, if CR-wave scattering is inefficient, CRs are less confined by Alfv\'en waves and can diffuse more freely along magnetic field lines. The transport of CRs becomes primarily diffusive, which we characterise by the CR diffusion coefficient $\kcr \sim \nu^{-1}_\mathrm{cr}$. In this regime, CRs experience reduced wave-particle interactions resulting in an increased effective CR transport speed, $\varv_\mathrm{cr}$, which redistributes them faster throughout the CGM. 

Cosmic rays transfer part of their energy directly to the surrounding gas through hadronic and Coulomb interactions \citep{Pfrommer2017}. Additionally, there is an indirect channel for energy transfer mediated by Alfv\'en waves. These waves interact with the plasma and undergo non-linear Landau damping \citep{Miller1991} gradually transferring their energy to the gas. This damping reduces the wave amplitude, lowers the CR scattering rate, and also heats the surrounding gas. In this work, we focus exclusively on non-linear Landau damping, which is considered the dominant damping mechanism in CGM-like environments.

\begin{figure}[!ht]
    \centering
    \resizebox{\hsize}{!}{\includegraphics{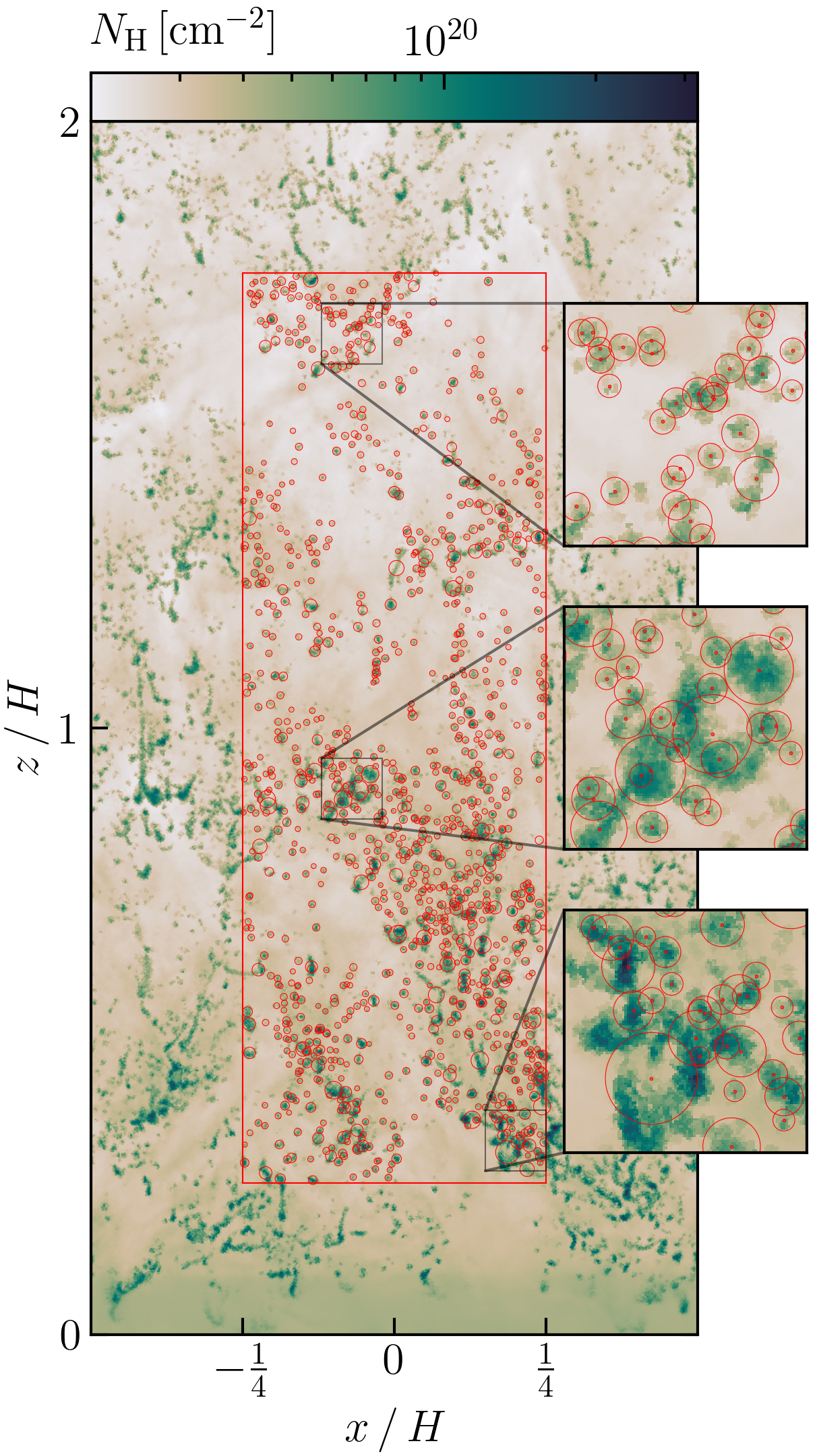}}
    \caption{Projection in $x$-$z$ of the hydrogen column density, $\NH$. Red circles represent the radii of clouds detected by the cloud-identification algorithm and red dots indicate their respective centroids. We note that overlapping circles result from the projection of clouds located at different $y$ positions and do not necessarily represent physical overlaps in three-dimensional space. }
    \label{fig:cloudfinder}
\end{figure}

\subsection{Cold gas metrics}
We utilised four key metrics to assess the results of our simulations: density fluctuations, cold mass fraction, cold mass volume, and cold mass flux. Each parameter offers valuable insights into different aspects of the underlying physical processes.

The density fluctuations,
\begin{equation}
    \label{eq:densitycontrast}
    \frac{\delta\rho}{\rho} = \frac{\rho - \Bar{\rho}}{\Bar{\rho}},
\end{equation}
quantify local perturbations in gas mass density, $\rho$, relative to the global volume-weighted average background density, $\Bar{\rho} = \sum_i (\rho_i V_i) / \sum_i V_i$, where $V_i$ is the volume of the $i$th computational cell.
This metric helped us identify regions where the local density significantly deviates from the background, which is crucial for understanding the dynamics of structure formation and collapse. 

The cold mass fraction,
\begin{equation}
    f_m = \frac{m_\mathrm{cold}}{M},
\end{equation}
represents the proportion of cold gas mass, $m_\mathrm{cold}$, relative to the total mass, $M$, within the analysed domain. This metric provides insights into the efficiency of gas cooling and conversion from the hot into the cold phase by TI.

The cold mass volume,
\begin{equation}
    f_V = \frac{V_\mathrm{cold}}{V},
\end{equation}
describes the spatial volume occupied by cold gas, $V_\mathrm{cold}$, relative to the total volume, $V$, offering a measure of the extent and distribution of cooler and denser regions. This parameter provides information about the size and mean density of collapsing regions.

The cold mass flux,
\begin{equation}
    \label{eq:coldmassflux}
    \frac{\Dot{m}_\mathrm{cold}}{\Dot{m}_\mathrm{ff}} = \frac{\rho_\mathrm{cold}}{\rho_0} \frac{\varv_\mathrm{cold}}{\varv_\mathrm{ff}},
\end{equation}
characterises the rate at which cold gas is accreting towards the midplane. Here, $\rho_\mathrm{cold}$ is the density of the cold gas, and $\varv_\mathrm{cold}$ denotes the velocity of infalling cold material. The reference values $\rho_0$ and $\varv_\mathrm{ff}=H/\tff$ are normalisation factors where $\rho_0$ is the initial density in the centre of the simulation domain, $\varv_\mathrm{ff}$ is a velocity defined by the scale height $H$, and the free-fall time $\tff$.

For the purpose of our analysis, a gas parcel is classified as cold if its temperature is at most $5\times10^{4}\,\unitT$. To mitigate contamination from potential boundary effects, we assess these metrics within a specified subrange of the computational domain, bounded by $0.25H\leq |z| \leq 2.75H$. This range is chosen to avoid unphysical behaviours near the simulation boundaries that could artificially influence the analysis. Within this region, we calculate each metric for every snapshot of the simulation, enabling a consistent comparison across time and different simulation setups. We run each simulation for $8\,\tcool$, using a fiducial value of $\tau=\tcool/\tff=0.3$. Typically, we restrict our analysis up to $t=7\,\tcool \approx 2\,\tff$ as the system's dynamics are expected to be significantly altered beyond this period.

\subsection{Cold cloud identification}
\label{subsec:coldcloudidentification}
To identify individual cold clouds in our simulations during post-processing, we utilised a watershed-like segmentation algorithm that leverages the Voronoi tessellation inherent to \textsc{Arepo}'s computational mesh.
The process begins by reconstructing the Delaunay triangulation \citep{SciPy} based on the mesh-generating points of the snapshot. In this triangulation, nodes correspond to the geometric centres of the original Voronoi cells, and edges represent connections between neighbouring cells. 

For each node, we determined and stored the indices of all its neighbouring nodes.
The nodes that satisfy specific temperature and density criteria, $T_\mathrm{node} \leq T_\mathrm{thresh}$ and $\rho_\mathrm{node} \geq \rho_\mathrm{thresh}$, were flagged as cloud candidates. From these candidates, spatially connected groups of nodes were identified using a graph-based clustering approach \citep{Csardi2005}. Any connected group containing at least five candidates was classified as a `cloud.' By construction, these clouds are disjointed and non-overlapping.
We estimated the effective radius of each cloud by first calculating the total volume of all associated cells. Assuming a spherical cloud geometry, the cloud radius was calculated by 
$r_\mathrm{cloud} = \left( 3 V_\mathrm{cloud} / 4\pi \right)^{1/3}$,
where $V_\mathrm{cloud}$ is the total volume of the cloud's cells. The position of each cloud was determined by its centroid, which we calculated as the mass-weighted average of the positions of all computational cells associated with the cloud. 
An example of the cloud-identification algorithm is illustrated in Fig.~\ref{fig:cloudfinder}.

\section{Simulation setup}
\label{sec:numericalsetup}
\subsection{Numerical framework}
We performed 3D CRMHD simulations with the moving-mesh code \textsc{Arepo} \citep{Springel2010, Pakmor2016} using standard parameters for mesh regularisation \citep{Vogelsberger2012, Weinberger2020}, and the magnetic divergence cleaning method described in \citet{Powell1999} using the implementation of \citet{Pakmor2013}. The equations of the two-moment CRMHD formalism as described in \citet{Thomas2019} are implemented using a finite-volume method \citep{Thomas2021}. In this approach, the CRMHD equations apply the P1 Eddington approximation for closure, which yields stable and consistent outcomes when compared to closure schemes of higher orders \citep{Thomas2022}.

The \textsc{Arepo} code provides the ability to refine and de-refine computational cells based on arbitrary user-defined criteria. We used the standard Lagrangian refinement scheme of \textsc{Arepo} to keep the mass enclosed by a computational cell nearly constant. If the mass of a cell deviates from its target value\footnote{Our fiducial run has a target gas mass of $m_\mathrm{target}=318\,\mathrm{M}_\odot$.} by more than a factor of two, the cell is refined or de-refined until the criterion is satisfied again. On top of that, in order to avoid significant volume deviations and thus large density gradients between neighbouring cells, computational cells are refined if one of their Voronoi neighbours has a volume that is smaller by a factor of 5. To manage computational resources efficiently, we set the minimum volume for a computational cell to $\left(H/2N\right)^3$, where $N$ represents the number of cells\footnote{Our fiducial run has $N=128$.} used to resolve one scale height, $H$, in the initial mesh.

Cosmic ray transport is computed assuming a reduced speed of light, $c_\mathrm{red}=3000~\mathrm{km\,s^{-1}}$. We subcycle the CR transport along the magnetic field lines ten times for each call of the MHD solver. This choice ensures that the characteristic speeds of sound and Alfv\'en waves, which typically peak at $\sim400~\mathrm{km\,s^{-1}}$, are well separated from the reduced speed of light, which is thus still the fastest signal speed in the simulation. We implement CR cooling following the approach by \citet{Pfrommer2017} such that the effective CR transport velocity, $\fcr\,/(\gcr\ecr)$, remains constant during the CR cooling process. We note that in the low-density environment of our simulations ($n \sim 10^{-3}\,\mathrm{cm^{-3}}$), CR cooling has a minimal impact\footnote{In our fiducial run with $\Xcrinit=3$ and two-moment CR transport, the total CR energy decreases by $\lesssim 8$\% over the course of the simulation.} because the simulation time is short ($\sim 1$ Gyr) compared to the CR cooling time ($\sim 10$ Gyr) as shown by \citet{Ensslin2011}.

\subsection{Gravity}
Although gravity fundamentally influences TI \citep{Donahue2022}, the exact form of the gravitational potential is not critical for the subsequent evolution \citep{Choudhury2016}. We integrated a simple vertical gravity profile as proposed by \citet{McCourt2012}:
\begin{equation}
    \label{eq:gravity}
    \bs{g}(z) = g_0 \frac{z/a}{\sqrt{1+\left(z/a\right)^2}} \hat{\bs{e}}_{z} = g(z) \, \hat{\bs{e}}_{z},
\end{equation}
where $z$ is the vertical distance from the midplane, and $a=H/10$ serves as a softening parameter for gravity that ensures that the profile transitions smoothly to zero near the centre of the simulation box, that is, for $|z|<a$. At large radii, the gravitational acceleration reaches a nearly constant value, determined by $g_0 = c_\mathrm{iso}^2/H$, where $c_\mathrm{iso}$ is the initial isothermal sound speed at the midplane. The corresponding free-fall time for a gas parcel at altitude $z$ is
\begin{equation}
    \label{eq:freefalltime}
    \tff = \sqrt{ \frac{2z}{g(z)} }.
\end{equation}
In our simulations, we do not account for self-gravity of the gas. This simplification is justified in CGM-like, low-density environments where the characteristic timescale for gravitational collapse of a self-gravitating gas cloud, $t \simeq (\mathrm{G}\rho)^{-1/2}$, considerably exceeds the timescale for the collapse due to compression induced by TI. Self-gravity becomes important only if the mass of the initial density perturbation exceeds the Jeans mass \citep{Li2020}.

\subsection{Cooling and heating}
To establish a thermodynamic environment consistent with the conditions in the CGM, we use a realistic cooling function that is designed to model the relevant cooling processes. We account for three main cooling channels: Lyman-$\alpha$ cooling by \element{H} and \element{He} atoms, cooling from various metal lines, and bremsstrahlung cooling due to free-free emission. We capture these processes in the collision-ionisation equilibrium (CIE) approximation and compile a look-up table with the resulting cooling rates. This table is then used to calculate the cooling rate in the simulation. To model the Lyman-$\alpha$ cooling process, we solve the ionisation equilibrium of \element{H} and \element{He} following \citet{1992Cen} and account for the ionisation due to an extragalactic ultraviolet background (UVB). We use the redshift $z=0$ UVB of \citet{2019Puchwein} and the ionisation cross-sections of \citet{1996Verner} to calculate the respective ionisation rates of H and He. The Ly$\alpha$ cooling rates are taken from \citet{1992Cen}. For metal line cooling, we calculate CIE cooling rates using \textsc{Chianti} \citep{1997Dere} for the elements \element{C}, \element{N}, \element{O}, \element{Si}, \element{Ne}, \element{Fe}, and \element{Mg} assuming solar metallicity using the abundances of \citet{2009Asplund}. These atoms are selected because they are the strongest metal coolants. Cooling by bremsstrahlung due to free-free emission from free electrons is calculated using the description of \citet{2018Ziegler}. The effective cooling rate in our simulation can be formulated as
\begin{equation}
    \label{eq:gascooling}
    \mathcal{C} = \nH^2\,\Lambda(\nH, T),
\end{equation}
where the volumetric cooling function, $\Lambda(\nH, T)$, compiles all the information regarding radiative cooling and heating processes outlined above. We introduce a cut-off temperature for the cooling function at $T_\mathrm{cut}=10^4\,\unitT$, which is achieved by employing a $\tanh$ profile that causes the cooling function to smoothly transition to zero at around $T_\mathrm{cut}$ across a temperature range of $\Delta T=100$~K. Gas with temperatures below $T_\mathrm{cut}$ remains in a cool state and undergoes no reheating due to any physically relevant processes.

Observational data suggest that the CGM is close to a state of global equilibrium \citep{Werk2014, Miller2015}. However, this balance is likely dynamic \citep{Tumlinson2011} rather than static. These findings imply a global heating mechanism that prevents the CGM from a cooling catastrophe. Although, the exact nature of the heating process remains uncertain, \citet{Zhuravleva2014} validated its plausibility by demonstrating that turbulent heating can offset radiative cooling across a range of radii in both the Perseus and Virgo clusters. Their work provided observational evidence that turbulent dissipation may serve as a significant energy source, potentially stabilising the intracluster medium against runaway cooling. This mechanism is also relevant on galactic scales, where turbulence, driven by SN feedback, AGN outflows, and cosmic inflows, plays a similar role in balancing radiative losses \citep{Voit2018, Buie2020, Faucher-Giguere2023}.
To address these observations, we introduce a heating mechanism that counterbalances cooling globally while permitting deviations from thermal equilibrium locally \citep{McCourt2012}. We subdivided the vertical extent of the simulation box into $N/2$ bins of equal height; collected the energy lost by radiative cooling in each of these bins at every time step,
\begin{equation}
    \Delta E_\mathrm{th,\,bin} = \sum m_i \, \Delta u_i \, ;
\end{equation}
and redistributed this energy inside the same bin by applying a mass-weighted scheme:
\begin{equation}
    \Delta E_{\mathrm{th}, i} = \frac{m_i}{M_\mathrm{bin}} \Delta E_\mathrm{th,\,bin} \, .
\end{equation}
Here, $u_i$ denotes the internal specific energy, and $m_i$ is the mass of cell $i$. The term $M_\mathrm{bin}$ is the cumulative mass of the bin containing the cell $i$.

\begin{figure}
    \centering
    \resizebox{\hsize}{!}{\includegraphics{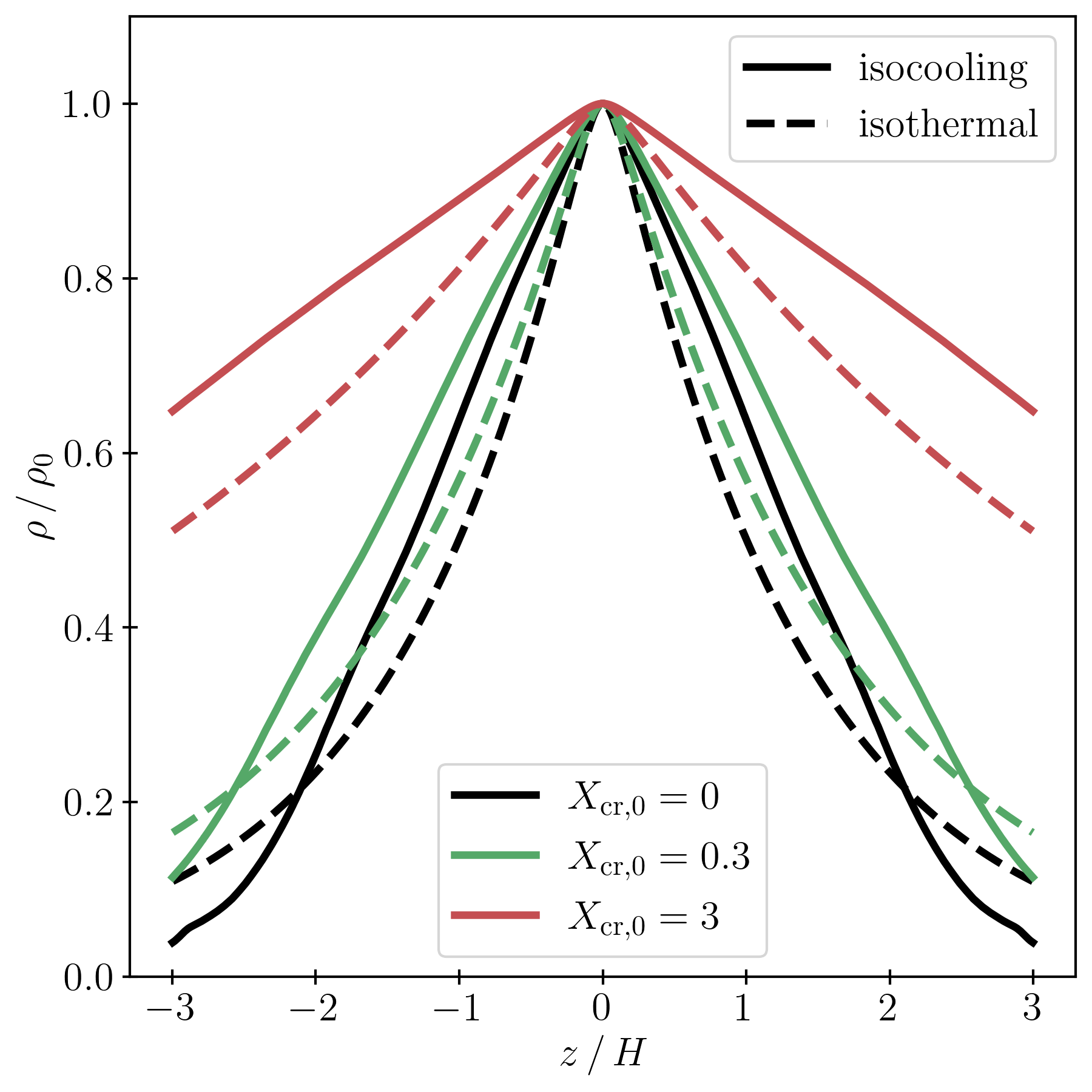}}
    \caption{Initial density profiles of the simulation box for various $\Xcrinit$. The solid lines show the implemented isocooling density profile, the dashed lines show the corresponding isothermal density profile with $T_0=10^6\,\unitT$ for comparison. For a higher initial $\Xcrinit$, more mass is contained in the atmosphere because the additional CR pressure supports the HSE without contributing to the gravitational force.}%
    \label{fig:densityprofiles}
\end{figure}

\subsection{Initial conditions}
The simulation domain models a gas column symmetrically extending from the ``galactic midplane'' far into the CGM. The computational box has the dimensions $1H\times1H\times6H$ along the $x$, $y$, and $z$ axes, where $H=30.11$ kpc in our fiducial run. We place the origin of the coordinate system in the centre of the box (cf.\ Fig.~\ref{fig:setup}). The $x$ and $y$ boundaries are periodic while the boundaries in $z$-direction allow gas to exit the computational domain. Here, $H$ denotes the characteristic scale height of the plasma, determined such that the ratio of cooling time to free-fall time, $\tau=\tcool/\tff$, is equal to the desired value at $z=\pm H$. The initial grid is uniform with a fiducial resolution of $128 \times 128 \times 768$ gas cells. To assess the convergence of our simulations, we conduct a subset of runs at resolutions corresponding to a quarter, half, and twice the fiducial value (see Table~\ref{tab:simulations}).

\paragraph{Density profile}
The initial gas distribution in each simulation follows an isocooling stratification \citep{Butsky2020}, implying the initial cooling time to be constant throughout the domain:
\begin{equation}
    \label{eq:coolingtime}
    \tcool = \frac{\varepsilon_\mathrm{th}}{\nH^2 \Lambda(\nH, T)} = \mathrm{const.}
\end{equation}
Combining this condition with the assumption of hydrostatic equilibrium (HSE), $\mathrm{d}P_\mathrm{tot} / \mathrm{d}z = -\rho(z)\,g(z)$, we derived the differential temperature profile for an isocooling atmosphere:
\begin{equation}
    \label{eq:temperatureprofile}
    \frac{\mathrm{d} T}{\mathrm{d} z} = \frac{\mu m_\mathrm{p}}{\eta k_\mathrm{B}} g(z) \left( 1 + \frac{1 - \Lambda_{T}}{1 + \Lambda_{\nH}}\right)^{-1},
\end{equation}
where $\mu$ is the mean molecular weight, $\mP$ denotes the proton mass, and $\kB$ is Boltzmann's constant. The quantities $\Lambda_{T} = \partial \ln{\Lambda} / \partial \ln{T}$ and $\Lambda_{\nH} = \partial \ln{\Lambda} / \partial \ln{\nH}$ describe the logarithmic derivatives of the cooling function with respect to temperature and density, respectively. $\eta = 1 + \Xcrinit + \Xmaginit + \Xkininit$ accounts for the contributions of thermal, CR, magnetic, and kinetic pressure to the total pressure, $P_\mathrm{tot}=\eta \Pth$.
We integrate Eq.~\eqref{eq:temperatureprofile} to obtain the vertical temperature profile and use the condition in Eq.~\eqref{eq:coolingtime} to calculate the corresponding pressure profile for an isocooling atmosphere in HSE. We present some of these initial density profiles for varying $\Xcrinit$ in Fig.~\ref{fig:densityprofiles}. 

\begin{table}[!t]
\centering
\caption{Parameters of all of our simulations.}
\label{tab:simulations}
\begin{tabular}{c c c c c l}
    $\Xcrinit$ & $\varv_\mathrm{cr,0}$ & $\kcr$ & $\tau$ & $H$ & Resolution \\ \hline\hline
    \multicolumn{6}{c}{Advective CRs} \\
    $0.03, 0.3, 3$ & $-$                & $-$               & $0.3$ & $30.11$ & fiducial\\ \hline
    \multicolumn{6}{c}{Diffusive CRs} \\
    $0.03, 0.3, 3, 30$ & $0$                & $3\times10^{27}$  & $0.3$ & $30.11$ & fiducial \\
    $0.03, 0.3, 3, 30$ & $0$                & $3\times10^{28}$  & $0.3$ & $30.11$ & fiducial \\
    $0.03, 0.3, 3, 30$ & $0$                & $3\times10^{29}$  & $0.3$ & $30.11$ & fiducial \\ \hline
    \multicolumn{6}{c}{two-moment CRs} \\
    $0.03, 0.3, 3, 30$ & $\vA$ & $-$               & $0.3$ & $30.11$ & fiducial \\
    $3$    & $\vA$ & $-$               & $0.3$ & $30.11$ & fiducial $/\,4$ \\
    $3$    & $\vA$ & $-$               & $0.3$ & $30.11$ & fiducial $/\,2$ \\
    $3$    & $\vA$ & $-$               & $0.3$ & $30.11$ & fiducial $\times\,2$\\ \hline
    \multicolumn{6}{c}{No CRs} \\
    $-$    & $-$                & $-$               & $0.1$ & $90.33$ & fiducial \\
    $-$    & $-$                & $-$               & $0.3$ & $30.11$ & fiducial \\
    $-$    & $-$                & $-$               & $0.3$ & $30.11$ & fiducial $\times\,2$ \\
    $-$    & $-$                & $-$               & $1$   & $9.033$ & fiducial \\
    $-$    & $-$                & $-$               & $3$   & $3.011$ & fiducial \\
    $-$    & $-$                & $-$               & $10$   & $1.004$ & fiducial \\
\end{tabular}
\tablefoot{
Additionally, all runs are initialised with $\Xmaginit=0.01$ and $\Xkininit=0.3$. Each two-moment CRMHD run is initialised with $\varepsilon_\mathrm{a,0}=10^{-3}\varepsilon_\mathrm{cr,0}$. The fiducial resolution starts with $128\times128\times768$ computational cells that are initially equally spaced. $\kcr$ is given in $\unitKappa$ and $H$ is given in kpc.
}
\end{table}

\paragraph{Velocity field}
The onset of TI requires the presence of local density perturbations. Previous studies have typically introduced such fluctuations by embedding isobaric density variations within the HSE \citep{McCourt2012, Butsky2020}. In contrast, our approach is based on observational evidence of turbulence in the CGM, as indicated by the broadening of absorption lines in quasar spectra, which suggests gas cloud motions \citep{Tumlinson2013}, and by \ion{Mg}{II} absorption systems, which further support the presence of turbulence in the CGM \citep{Huang2016}. Turbulent motions generate converging flows that change the local density of the plasma on short timescales. To more accurately model the formation of thermally unstable regions, we initialise our simulations with a turbulent velocity field that follows a Kolmogorov spectrum \citep{Kolmogorov1941} within the range $4 \leq k\,H / 2\pi \leq 256$, which transitions to white noise on larger scales. We adjust this velocity field to ensure a kinetic-to-thermal pressure ratio of $\Xkin=\Pkin/\Pth=0.3$ at every height. In our fiducial CR-dominated setup (described below), this method yields a velocity dispersion of approximately $\sigma_\varv \sim 33\,\mathrm{km\,s^{-1}}$ near the centre and $\sigma_\varv \sim 30\,\mathrm{km\,s^{-1}}$ at the edges of the simulation box.

\paragraph{Magnetic field}
Turbulence amplifies the magnetic field predominantly via the small-scale dynamo mechanism \citep{Brandenburg2005, Pakmor2017, Pfrommer2022} and determines the orientation of the magnetic field lines. To encapsulate the dynamics between turbulent gas motion and magnetic fields in our simulations, we introduce a turbulent magnetic field that mirrors the scaling of the previously mentioned velocity field. Recent numerical investigations of the CGM in cosmological galaxies by \citet{Pakmor2020} have shown an approximately constant relative magnetic pressure, $\Xmag=\Pmag/\Pth$, within the virial region. Thus, we vary the magnetic field strength in height following the trend of the thermal pressure profile to replicate this simulated trend. A Helmholtz decomposition is employed to ensure the magnetic field's divergence-free nature by eliminating its compressive component. This procedure yields a purely solenoidal magnetic field, which is subsequently normalised by a constant factor to ensure an average relative magnetic pressure of $\Xmag=0.01$ throughout the simulation domain.

\begin{figure*}[!ht]
    \centering
    \includegraphics[width=17cm]{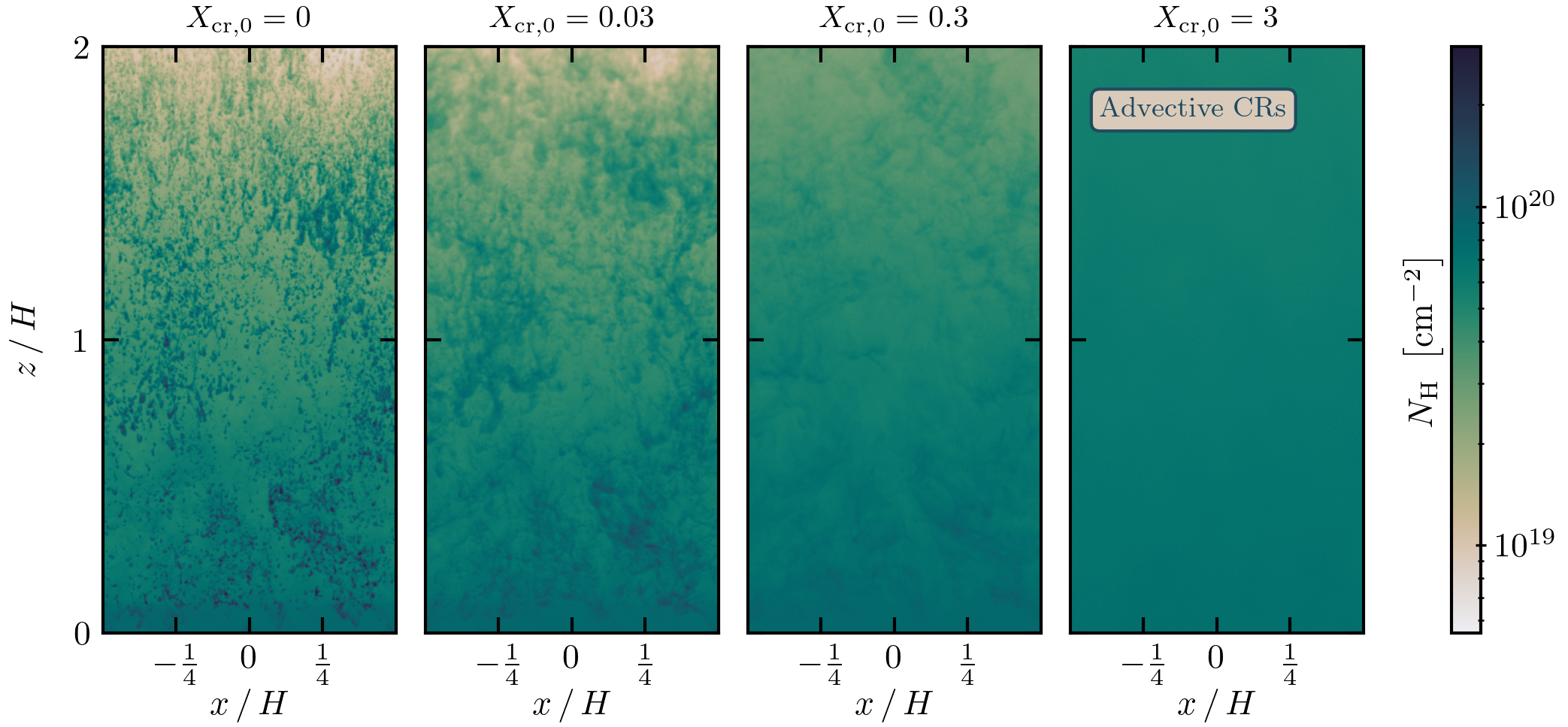}
    \caption{Projections of the hydrogen column density, $\NH$, employing purely advective CR transport. The panels show the $x-z$ plane of a section extending from $z=0$ to $z=2H$ and the depth of the projection is $1\,H$. The snapshot is at $t=7\,\tcool \approx 2\,\tff$. All simulations were initialised with different relative CR pressure, $\Xcrinit$. The density of the clouds decreases with increasing CR pressure, while the cloud size increases. In the CR pressure dominated run with $\Xcrinit=3$, cold gas condensation is entirely suppressed. This illustrates that advective CRs alter the morphology of the CGM significantly and suppress the formation of cold clouds.
    }%
    \label{fig:cradv}
\end{figure*}

\begin{figure*}
    \centering
    \includegraphics[width=17cm]{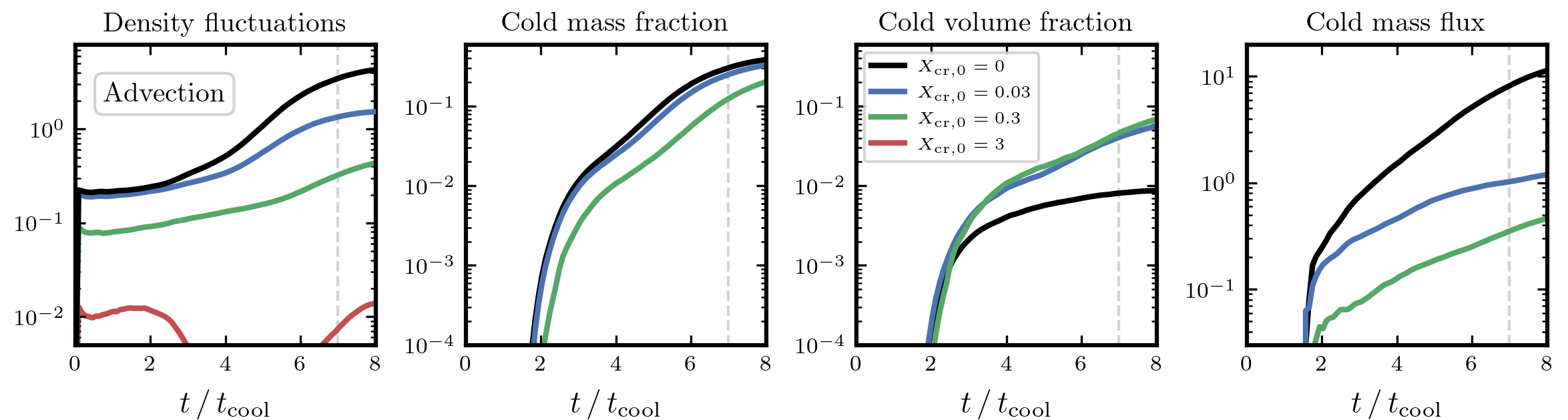}
    \caption{Time evolution of cold gas metrics for simulations employing advective CR transport with varying $\Xcrinit$. From left to right, we show density fluctuations, cold mass fraction, cold volume fraction, and cold mass flux. We evaluated each computational cell in the region $0.25\,H \leq |z| \leq 2.75\,H$. The dashed vertical line marks the simulation time corresponding to the results shown in Fig.~\ref{fig:cradv}.}
    \label{fig:cradv_coldmass}
\end{figure*}

\paragraph{Cosmic rays}
We include CR pressure, $\Pcr$, in the initial conditions as a constant fraction of the gas pressure, $\Pth$, in every computational cell, characterised by the parameter $\Xcr=\Pcr/\Pth$. Introducing this additional non-thermal pressure modifies the isocooling density profile compared to a setup without CRs. Specifically, the non-thermal pressure gradient necessitates an adjustment in the gravitational force to maintain equilibrium, resulting in an atmosphere with a higher mass content.
Figure~\ref{fig:densityprofiles} shows the resulting initial density profiles for various choices of $\Xcrinit$. We explore a range of initial values for $\Xcrinit$ spanning $0$, $0.03$, $0.3$, $3$, and $30$ to characterise different CGM environments, ranging from purely thermal to slightly CR-supported to strongly CR-dominated atmospheres. For runs modelling CR transport by pure diffusion, we vary the constant CR diffusion coefficient, $\kappa_0$, to simulate slow ($3\times10^{27}\,\unitKappa$), moderate ($3\times10^{28}\,\unitKappa$), and fast ($3\times10^{29}\,\unitKappa$) diffusion rates. In simulations employing the two-moment approach for CR transport, CRs initially stream along the magnetic field at the Alfv\'en speed, $\vA$, characterised by the initial CR flux density, $\fcr=\vA(\ecr+\Pcr)$. We initialise the energy density of Alfv\'en waves with $\varepsilon_\mathrm{a}=10^{-3}\ecr$.

\paragraph{Numerical values}
In the midplane of our simulation box, we initialised the values for gas density and temperature with $\rho_0=2\times10^{-27}\,\unitRho$ and $T_0=10^6\,\unitT$, respectively. These parameters in combination with the isocooling condition yielded a global cooling time for the gas of $\tcool\approx108\,\mathrm{Myr}$. Given our fiducial ratio of cooling time to free-fall time, $\tau=0.3$, we derive a scale height of $H=30.11\,\mathrm{kpc}$. We summarise the parameters for the different simulations in Table~\ref{tab:simulations}.

\section{Suppression of thermal instability by advective cosmic rays}
\label{sec:advectivecrs}
We start our investigation of the impact of CRs on TI by examining purely advective CR transport, which represents the limiting case of inefficient anisotropic CR transport. In this scenario, CRs are exclusively advected with the thermal gas and have no motion relative to the medium. CR transport can be approximated by such advection models when the timescale for CR transport is significantly longer than the timescales governing other relevant physical processes. This situation is partially realised in environments where strong bulk gas flows are present, such as at the launching sites of galactic outflows or within supernova remnants \citep{Ruszkowski2023, Armillotta2024}. All simulations in this section are initialised with $\tau=0.3$, $\Xmaginit=0.01$, $\Xkininit=0.3$, and different values for $\Xcrinit$. We run each simulation until $t=8\,\tcool$.

Figure~\ref{fig:cradv} illustrates how the pressure support of advective CRs changes the morphology of cold gas in the CGM. We show projections in $y$-direction of the hydrogen column density, $\NH$, after the simulations run for $7\,\tcool$. From left to right, we systematically vary the initial CR pressure, $\Xcrinit$, ranging from purely thermal ($\Xcrinit=0$) to CR-pressure dominated ($\Xcrinit=3$) scenarios. As the gas cools, denser regions lose their thermal pressure support more rapidly than their surroundings. These denser gas parcels sink and undergo compression due to the rising ambient pressure of their vicinity. In the simulation without CRs, pressure equilibrium is approximately sustained due to the short cooling time ($\tau=0.3$). This leads to ongoing compression until the gas temperature reaches a new equilibrium. However, the presence of CR pressure changes this scenario considerably. Even a minor CR pressure support of $\Xcrinit=0.03$ clearly alters the morphology of the collapsing regions. The inability of CRs to escape the contracting clouds in this model leads to an elevated non-thermal pressure support that counteracts the collapse and, with an increasing contribution of CRs, gradually limits the extent to which the gas can be compressed. 

\begin{figure*}[!ht]
    \centering
    \includegraphics[width=17cm]{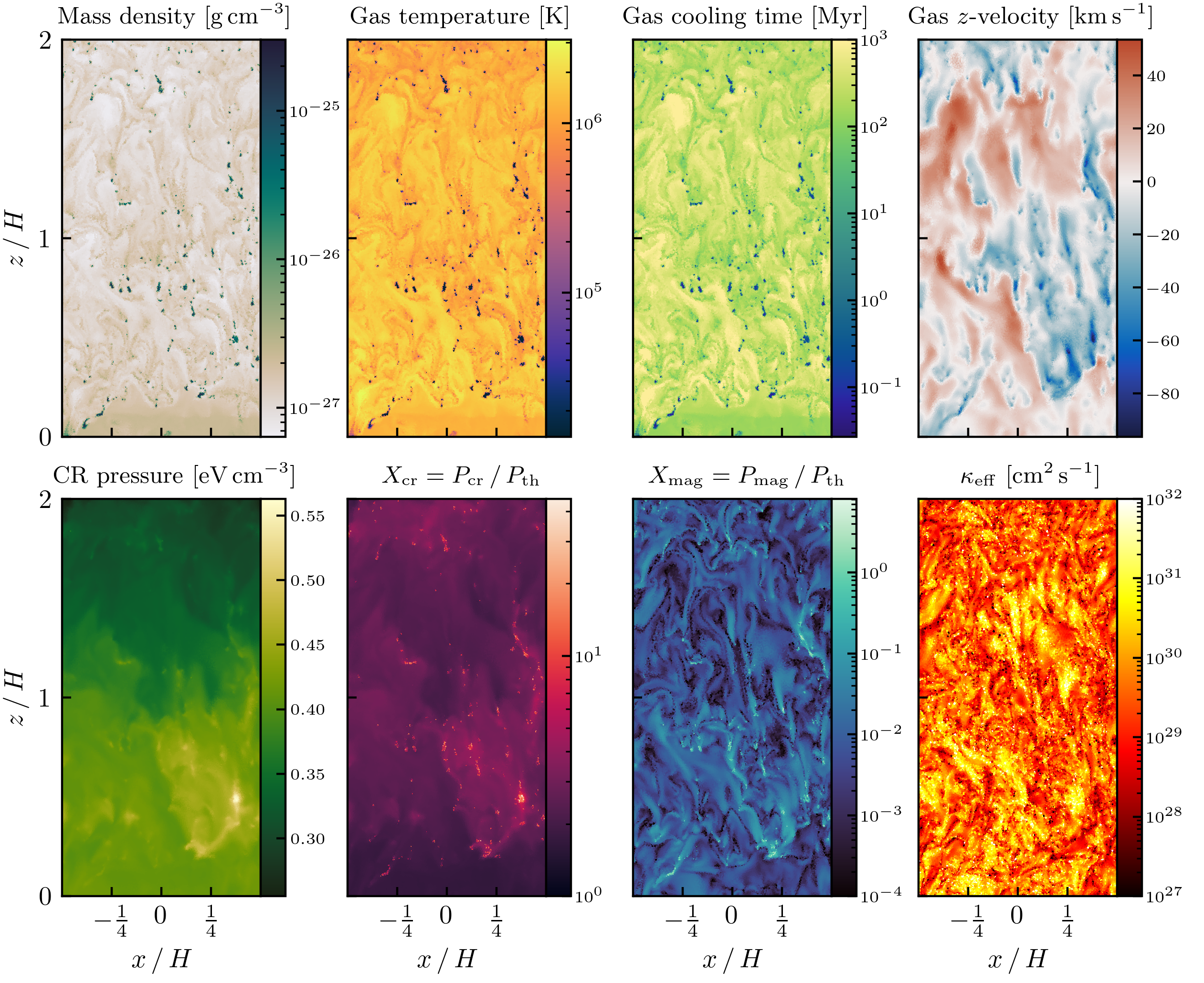}
    \caption{Gallery showing 2D slices of various quantities from the simulations employing two-moment CR transport with an initial CR pressure fraction of $\Xcrinit=3$. All snapshots are at $t=7\,\tcool$. The top row illustrates quantities of the thermal gas, while the bottom row presents variables related to CR transport.}%
    \label{fig:gallery}
\end{figure*}

The onset of TI is completely suppressed in the CR-pressure dominated run ($\Xcrinit=3$). While this suppression correlates with increased CR pressure, it is not the CR pressure alone that causes it. Instead, the suppression arises from the method used to generate the initial density perturbations required to trigger collapse. In all simulations, the initial ratio of kinetic to thermal pressure, $\Xkininit=P_\mathrm{kin,0}\,/\,P_\mathrm{th,0}$, is fixed to 0.3, regardless of the CR pressure. As $\Xcrinit$ increases, the thermal pressure makes up a smaller fraction of the total pressure, and thus the same $\Xkininit$ corresponds to a smaller absolute level of kinetic energy relative to the total pressure. This results in significantly weaker initial density perturbations compared to runs with lower $\Xcrinit$. These weaker perturbations are more easily smoothed out by turbulent heating introduced via the initial velocity field, further reducing fluctuations and ultimately suppressing the development of TI.

This behaviour is clearly illustrated in Fig.~\ref{fig:cradv_coldmass}, which presents (from left to right) the density fluctuations, cold mass fraction, cold volume fraction, and cold mass flux for simulations with purely advective CR transport. The left panel shows that the stabilising effect of CRs becomes apparent as the initial density fluctuations decrease with increasing $\Xcrinit$. Notably, the cold mass fraction converges to nearly identical values across all simulations permitting TI, as expected, since the cooling function depends solely on gas density and temperature and is independent of CR pressure. Once a region enters TI, the subsequent runaway cooling proceeds unaffected by CRs. However, the volume of cold gas increases by a factor of $10-20$ compared to the non-CRs run, which is driven by the additional CR pressure counteracting the compression of thermally unstable gas. This influence is further underscored by the cold mass flux analysis: the increased volume and reduced density of cold gas result in stronger buoyancy forces, which reduce the cold mass flux in simulations with higher CR pressure support.

In previous work, \citet{Butsky2020} found that the onset of TI is entirely independent of CR pressure, which appears to contradict our results. However, we highlight that our approach to generating density perturbations differs significantly from theirs. While \citet{Butsky2020} initialise their simulations with a hydrostatic atmosphere containing small, pre-seeded density inhomogeneities, we evolve these perturbations self-consistently from an initial turbulent velocity field. Each procedure performs best for the distinct aspects of TI analysis: the method of \citet{Butsky2020} optimally captures the linear growth of TI whereas our more dynamical approach aims to mimic the more realistic evolution of astrophysical systems subject to turbulence. 

However, the idealised scenario of purely advective CR transport only partially applies in realistic environments, where CRs usually can actively diffuse or stream along magnetic field lines. The transition from purely advective to active CR transport fundamentally alters the CR pressure distribution, reduces the stabilising effect of CRs, and reshapes the dynamics of TI. We address this topic in the following section.

\section{Revival of thermal instability by self-confined cosmic rays}
\label{sec:2momentcrs}

\begin{figure*}
    \centering
    \includegraphics[width=17cm]{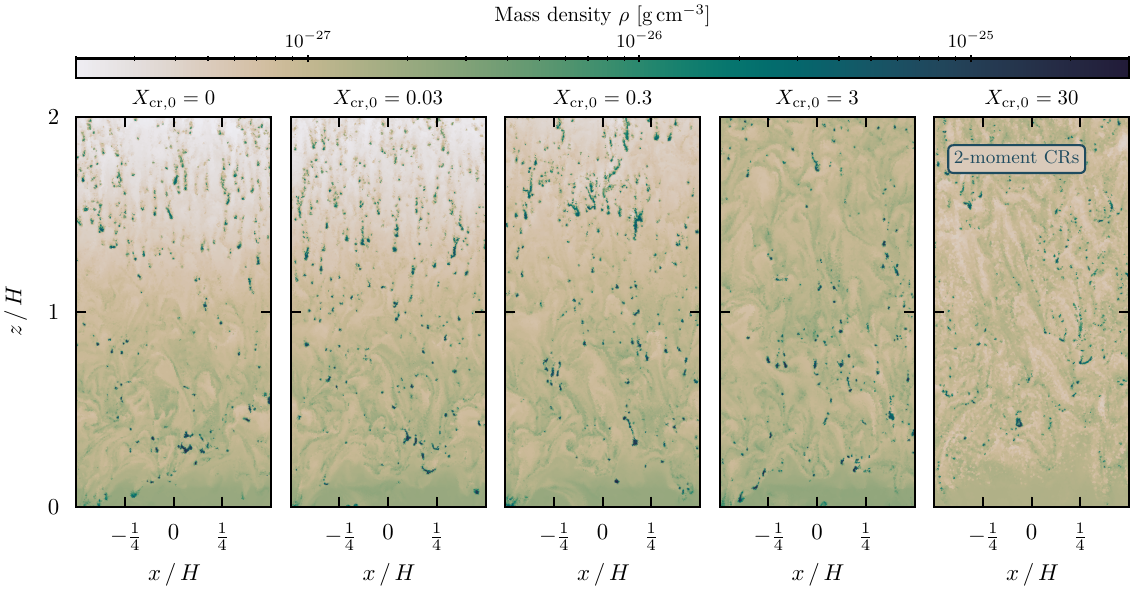}
    \caption{Slices showing the mass density in the $x-z$ plane centred at $y=0$ of simulations employing two-moment CR transport and different relative CR pressures, $\Xcrinit$. All snapshots are at $t=8\,\tcool$. The initial CR pressure barely influences the morphology of the cold gas in terms of cloud size or density.}%
    \label{fig:2momCRs}
\end{figure*}

\begin{figure*}
    \centering
    \includegraphics[width=17cm]{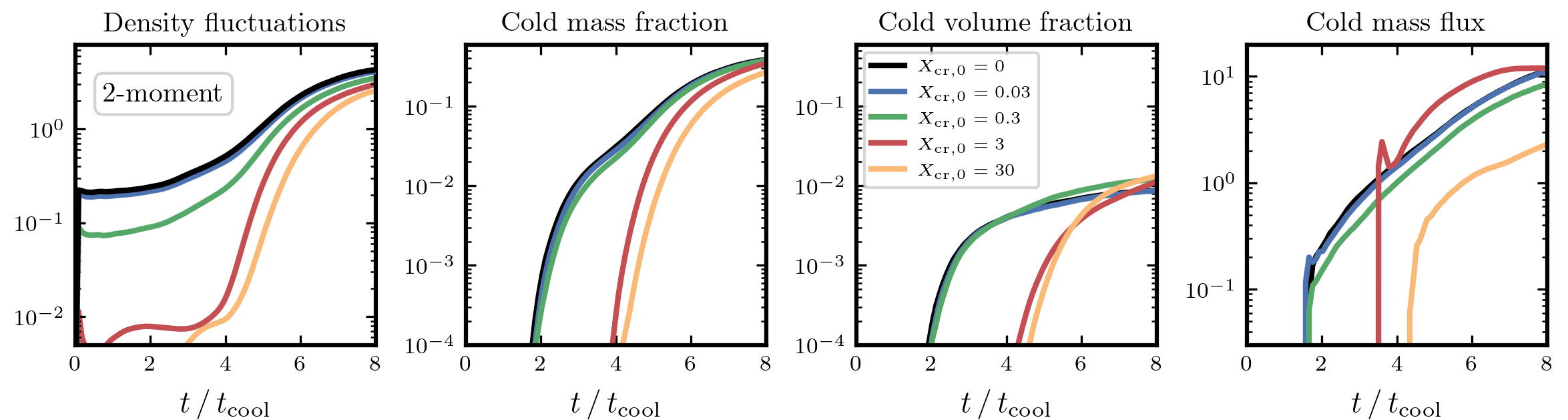}
    \caption{Time evolution of cold gas metrics for simulations employing two-moment CR transport with varying $\Xcrinit$. From left to right, we show density fluctuations, cold mass fraction, cold volume fraction, and cold mass flux. We evaluate each computational cell in the region $0.25\,H \leq |z| \leq 2.75\,H$. All cold gas metrics show a very similar evolution and are almost independent of the initial CR pressure.}%
    \label{fig:coldGasMetrics_2momCRs}
\end{figure*}

In realistic astrophysical environments, CRs interact with the surrounding medium in more complex ways than simply being advected with the gas. Active CR transport in the form of diffusion and streaming allows CRs to decouple from the gas and move along magnetic field lines at speeds that can exceed the local gas velocity. In the following, we discuss the implications of active CR transport on the formation of cold gas.

We begin our analysis with a general overview of our simulations. In Fig.~\ref{fig:gallery}, we present a gallery showcasing slices of various quantities from the simulation utilising two-moment CR transport with an initial relative CR pressure of $\Xcrinit=3$. The top row shows various gas quantities, namely density $\rho$, temperature $T$, cooling time $\tau_\mathrm{cool}$, and gas $z$-velocity $\varv_z$. The bottom row displays the CR pressure $\Pcr$, the relative CR pressure $\Xcr=\Pcr/\Pth$, the relative magnetic pressure $\Xmag=\Pmag/\Pth$, and the effective CR diffusion coefficient $\keff$ discussed below (see Sec.~\ref{sec:crdiffusioncoefficient}). This indicates that active CR transport induces substantial changes in the condensation process of cold gas compared to the advective CR scenario (cf.\ Fig.~\ref{fig:cradv} ). 

Starting from isocooling initial conditions, a two-phase medium emerges, characterised by a volume-filling, dilute hot phase ($\rho \lesssim 10^{-27}\,\unitRho$, $T \gtrsim 10^6\,\unitT$) and a dense cold phase ($\rho \gtrsim 10^{-25}\,\unitRho$, $T \sim 10^4\,\unitT$), the latter being composed of numerous small ($\sim 100\,\mathrm{pc}$) clouds with very short central cooling times ($\tcool \lesssim 100\,\mathrm{kyr}$). These overdense cold clouds fall towards the centre of gravity with high velocities so that the densest clumps reach infall speeds of $\gtrsim 90\,\mathrm{km\,s^{-1}}$. Interestingly, the upwards motions are caused by turbulence as well as the motions induced by our assumed mass-weighted heating that is globally offsetting radiative cooling. The CR pressure exhibits a relatively uniform distribution across the box (we point out the linear colour scale) with moderate peaks within and around the collapsing regions due to adiabatic compression of CRs. This compression effect also applies to the magnetic pressure, represented by $\Xmag$, which reaches values of about one to eight in and around the cold clouds as a result of the flux-frozen characteristic of the magnetic field. The range in $\Xmag$ is much larger in comparison to that of $\Xcr$, which is a direct consequence of fast CR transport out of the collapsing clumps. In fact, the effective CR diffusion coefficient spans an immense range of more than seven orders of magnitude with values between $\keff \sim 10^{27}\,\unitKappa$ and $\keff > 10^{32}\,\unitKappa$, with the largest values realised in the converging regions, which are also characterised by large values of $\Xmag$. 

Having identified substantial differences between outcomes under initially dominating CR pressurisation between the two-moment setup and the purely advective CR scenario, we extend our analysis to include a wider range of initial CR pressure values. To this end, we repeat the simulation suite from Sec.~\ref{sec:advectivecrs} and include a strongly CR-pressure-dominated atmosphere with $\Xcrinit = 30$. All simulations are initialised with $\tau=0.3$, $\Xmaginit=0.01$, $\Xkininit=0.3$, and are evolved for eight cooling times.

Figure~\ref{fig:2momCRs} demonstrates that under realistic conditions of active CR transport, variations in the initial CR pressure have minimal impact on the morphology of cold gas in the CGM. The figure shows gas mass density slices at $t=7\,\tcool$, with panels illustrating a systematic variation in $\Xcrinit$ from purely thermal conditions ($\Xcrinit=0$) to strongly CR pressure-dominated regimes ($\Xcrinit=30$). Remarkably, the number and size of condensing clouds remain nearly unchanged across all simulations, showing minimal dependence on the initial CR pressure. Even under the influence of large CR pressures ($\Xcrinit=30$), TI progresses without significant suppression. 

Figure~\ref{fig:coldGasMetrics_2momCRs} further supports these findings, showing the temporal evolution of cold gas metrics. As described in Sec.~\ref{sec:advectivecrs}, the initial density fluctuations decrease with increasing $\Xcrinit$ as a consequence of the reduced influence of the initial kinetic pressure, which is constant ($\Xkininit=0.3$) across all runs. This effect is particularly relevant in setups where the CR pressure is significantly higher than the kinetic pressure, that is, in simulations with $\Xcrinit \ge 3$. Consequently, cold gas forms delayed ($t \sim 4\,\tcool$) compared to thermally dominated runs ($t \sim 2\,\tcool$). After the simulations evolved for $t \sim 7\,\tcool$, nearly all metrics converge to similar values across the simulations. The only exception is the cold mass flux in the run with $\Xcrinit = 30$, which remains offset by approximately a factor of five in comparison to the other simulations. We attribute this effect to the exceptionally high CR pressure, which counteracts gravity and consequently reduces the infall velocity of the cold clouds.

Our analysis demonstrates that the amount of CRs has only a small impact (within a factor of two) on the final amount of cold gas or its volume filling fraction, see Fig.~\ref{fig:coldGasMetrics_2momCRs}. This finding is somewhat unexpected, because CR pressure is typically thought to play a significant role in shaping gas dynamics by contributing to additional non-thermal support and influencing cooling instabilities \citep{Butsky2020}. The lack of substantial morphological changes suggests that the interplay between CR transport, gas dynamics, and cooling is governed by other processes.
We attribute this outcome to the high CR transport speed, as reflected by the high values of the effective CR diffusion coefficient (see bottom right panel of Fig.~\ref{fig:gallery}), which enables CRs to escape collapsing regions before significantly influencing TI. In the following section, we explore this behaviour further by examining the effect of varying CR transport speeds, focusing specifically on changes in the effective CR diffusion coefficient, $\keff$, and its impact on the development of TI.

\begin{figure}
    \centering
    \resizebox{\hsize}{!}{\includegraphics{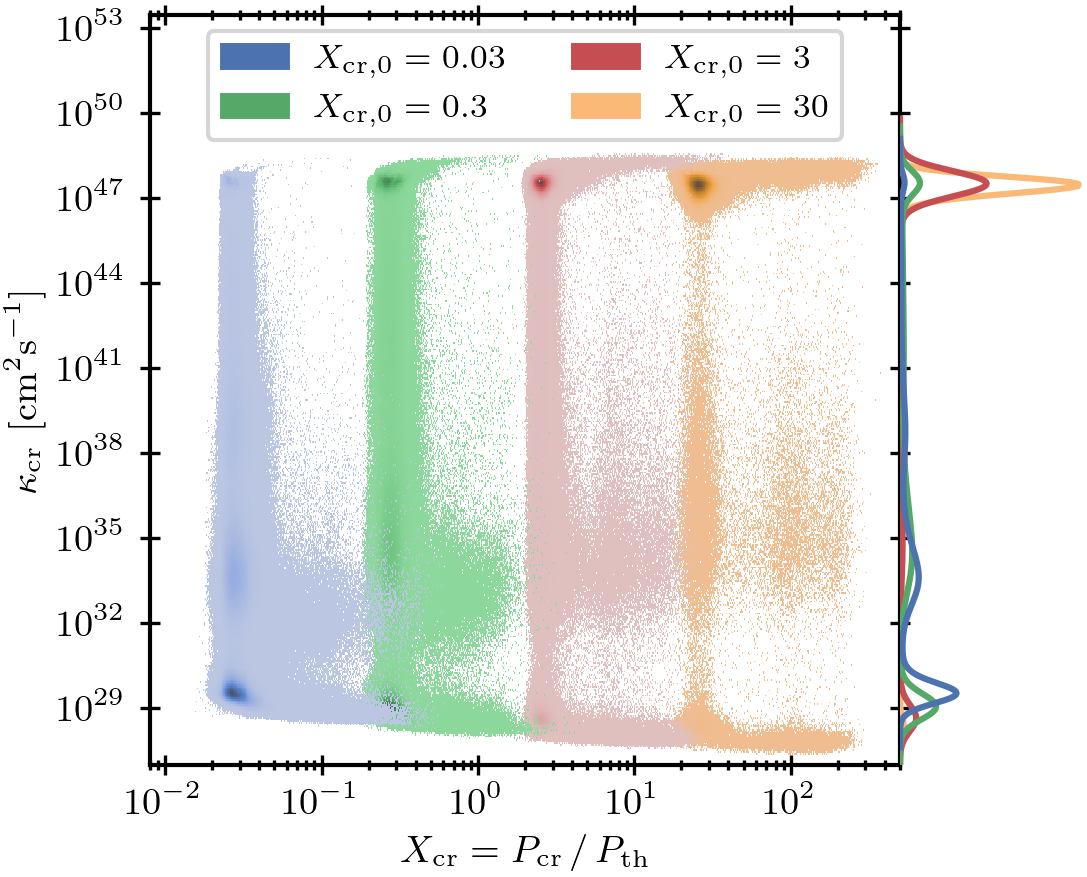}}
    \caption{Intrinsic CR diffusion coefficient, $\kcr$, for simulations utilising two-moment CR transport with varying initial values of $\Xcrinit$. 
    Scatter points represent the distribution of CRs as a function of $\kcr$ and $\Xcr$ at $t=6\,\tcool$, with darker colours indicating higher CR-energy-weighted probability density in a logarithmic colour scheme. The corresponding 1D probability density of $\kcr$ is displayed with a linear scaling on the right-hand side. In the regime of efficient CR scattering (small $\kcr$), larger $\Xcrinit$ values increase the overall CR energy density in the CGM allowing for more CR energy to be converted into gyroresonant Alfv\'en waves and larger CR scattering rates, thereby reducing $\kcr$ (see Eq.~\ref{eq:intrinsiccrdiffusioncoefficient}). In regions of a smooth CR distribution, the driving of Alfv\'en waves is inefficient in comparison to wave damping processes so that CRs are poorly coupled to the plasma, leading to a very large value of $\kcr$. }%
    \label{fig:kappa_intrinsic}
\end{figure}

\section{Impact of cosmic ray transport speed on thermal instability}
\label{sec:crtransportspeed}
The results from the previous section suggest that CRs escape too quickly from collapsing regions to meaningfully affect the condensation of cold gas. Therefore, in this section, we explore the significance of the CR transport speed and begin by analysing how CRs are transported in the presence of TI.

Relative to astrophysical plasmas, the CR distribution primarily propagates via streaming and diffusion along magnetic field lines. In the streaming regime, CRs travel at the local Alfv\'en speed. As they propagate, they excite Alfv\'en waves that pitch-angle scatter the CRs. In turn, these plasma waves lose their energy through wave damping processes. On the other hand, (anisotropic) diffusion describes CR transport along magnetic field lines down their pressure gradient. This can be described as a random walk with a single step representing a CR pitch-angle scattering event of off small-scale magnetic fluctuations. In 1-moment CR transport models, both processes are assumed to be in steady-state: in the ISM, the CR scattering rate is typically constant and independent of local physical conditions, producing ``canonical'' values of $\kcr \sim 3\times10^{27-28}\,\unitKappa$ \citep{Ruszkowski2023}, while perfect CR streaming assumes a tight coupling between CRs and gyroresonant Alfv\'en waves, which results in a streaming velocity equal to the Alfv\'en speed. However, these assumptions hold only in environments where Alfv\'en-wave generation and damping are perfectly balanced so that the energy in these waves is (i) constant and maintains a fixed scattering rate, and (ii) sufficient to continuously couple CR motion to the waves. In realistic environments, these two conditions are not permanently maintained. The generalisation to more realistic CR transport is provided by the two-moment CR transport approach.

\subsection{Cosmic ray transport in the two-moment picture}
\label{sec:crdiffusioncoefficient}
The CR transport in the framework of CRMHD is also characterised by a combination of streaming and diffusion. The CRs stream along magnetic field lines at the CR streaming speed \citep{Thomas2019}
\begin{equation}
    \varv_\mathrm{st} = \vA \frac{\nu_{+} - \nu_{-}}{\nu_{+} + \nu_{-}},
\end{equation}
where $\vA$ denotes the local Alfv\'en speed while $\nu_{+}$ and $\nu_{-}$ are the scattering rates of CRs with respect to co- and counter-propagating Alfv\'en waves, respectively. If one of these wave types is damped away and its corresponding scattering frequency vanishes, this equation approaches the limiting case of 1-moment streaming where CRs stream down their pressure gradient, $\varv_\mathrm{st} = - \vA\,\mathrm{sign}\left(\bs{b}\bcdot \bnabla \Pcr\right)$ \citep{Zweibel2013, Pfrommer2017, Thomas2023}.

When the transport velocity of CRs exceeds $\vA$, the excess velocity is attributed to CR diffusion relative to the Alfv\'en waves. This diffusive component is quantified by the intrinsic CR diffusion coefficient \citep{Thomas2019}:
\begin{equation}
    \kappa_\mathrm{\pm} = \frac{c^{2}}{3\nu_\mathrm{\pm}} = \frac{8}{9\pi} \frac{\gamma m c^3}{e B} \frac{B^2}{\varepsilon_\mathrm{a,\pm}},
    \label{eq:intrinsiccrdiffusioncoefficient}
\end{equation}
with its total value given by
\begin{equation}
    \kcr = \left(\kappa^{-1}_{+} + \kappa^{-1}_{-}\right)^{-1},
\end{equation}
where the $+$ and $-$ signs of the subscripts correspond to quantities relative to co-propagating and counter-propagating Alfv\'en waves, respectively. Here, $B$ is the magnetic field strength, $e$ is the elementary charge, $c$ is the speed of light, and $\gamma = 2$ represents the typical Lorentz factor of giga-electronvolt CRs with rest mass $m$. The energy density content of Alfv\'en waves, $\varepsilon_\mathrm{a}$, controls the level of CR diffusivity: in regions of high $\varepsilon_\mathrm{a}$, CRs scatter frequently and are tightly coupled to the Alfv\'en waves, reducing their effective diffusive transport. Conversely, in regimes with low $\varepsilon_\mathrm{a}$, CRs adopt higher diffusion coefficients, making their transport predominantly diffusive. 

\begin{figure}
    \centering
    \resizebox{\hsize}{!}{\includegraphics{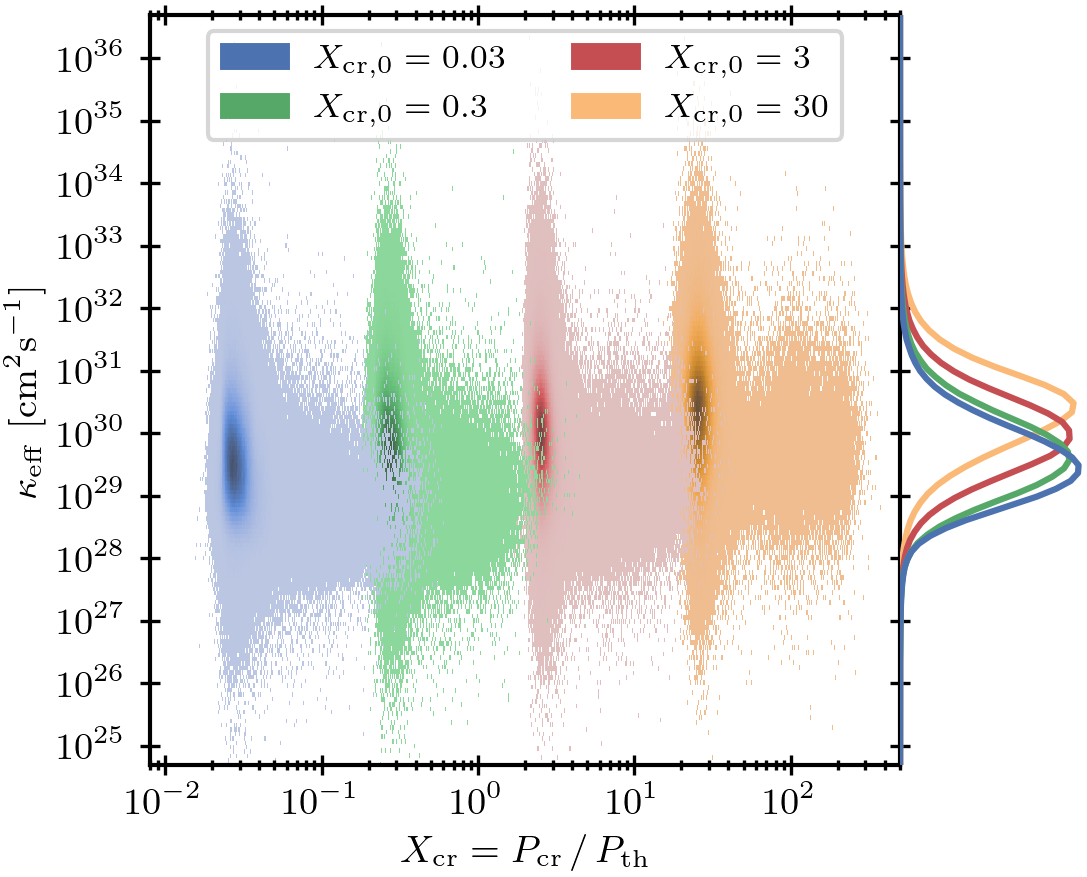}}
    \caption{Same as Fig.~\ref{fig:kappa_intrinsic} but for the effective CR diffusion coefficient, $\keff$.}%
    \label{fig:kappa_eff}
\end{figure}

\begin{figure*}
    \centering
    \includegraphics[width=17cm]{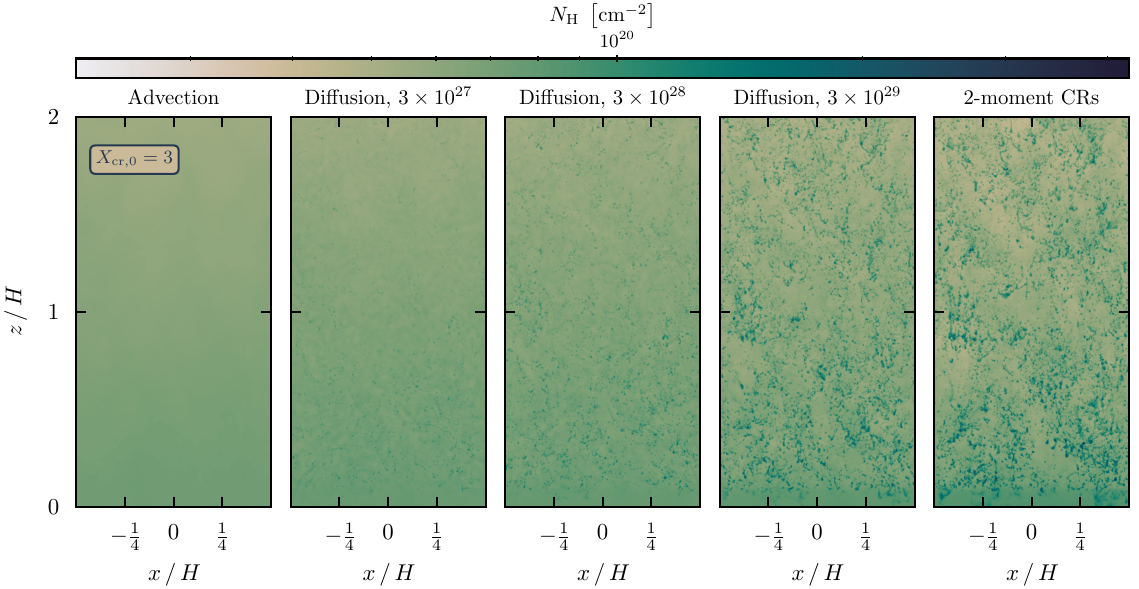}
    \caption{Projections of the hydrogen column density, $\NH$, for simulations utilising different CR transport models. All runs were initialised with $\Xcrinit=3$, and the snapshots are at $t=7\,\tcool$. The figure demonstrates that the onset of collapse is governed by the CR transport velocity rather than the CR pressure.}%
    \label{fig:crtransports}
\end{figure*}

\begin{figure*}
    \centering
    \includegraphics[width=17cm]{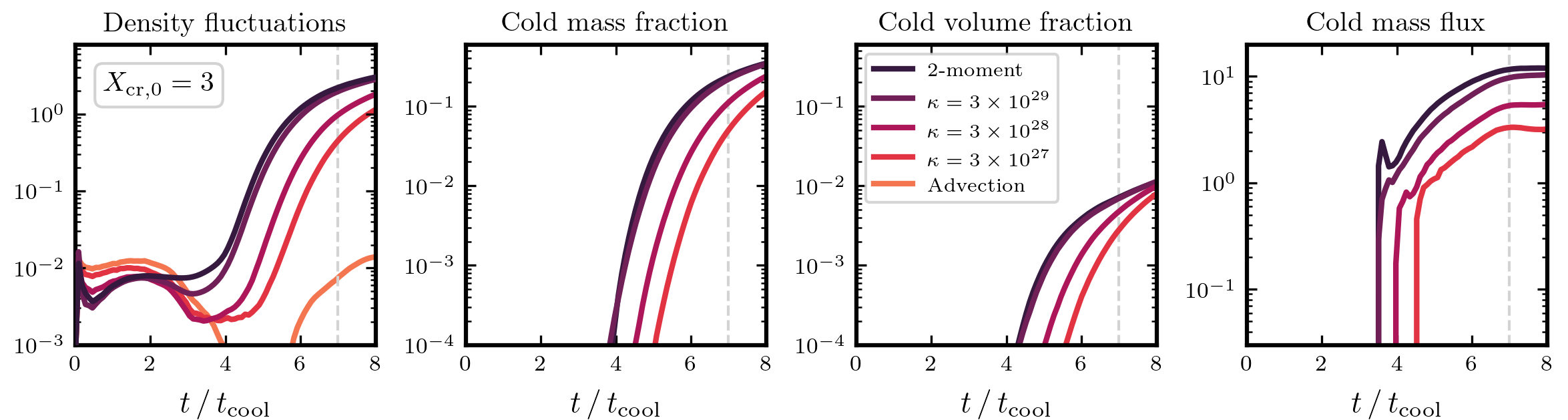}
    \caption{Time evolution of cold gas metrics for simulations with different CR transport models and $\Xcrinit=3$. Metrics are evaluated for computational cells within $0.25\,H \leq |z| \leq 2.75\,H$. The dashed vertical line marks the time corresponding to the snapshot shown in Fig.~\ref{fig:crtransports}. Lower CR transport velocities delay the onset of cloud collapse and reduce the amount of condensed cold gas.}%
    \label{fig:crtransport_coldmass}
\end{figure*}

In Fig.~\ref{fig:kappa_intrinsic}, we show the CR distribution in the total intrinsic CR diffusion coefficient, $\kcr$, and $\Xcr$ at $t=6\,\tcool$. Simulations that are initialised with different values of $\Xcrinit$ result in $\Xcr$ values that range from approximately 0.5 to 10 times their initial relative CR pressure. Two distinct trends emerge from the figure. First, the intrinsic diffusion coefficients span an extensive range, from the slow-diffusion regime with $\kcr \sim 2 \times 10^{27}\,\unitKappa$ to values exceeding $10^{47}\,\unitKappa$. We observe that the fraction of CRs in the high-diffusivity regime increases with increasing $\Xcrinit$ and even dominates the CR population in simulations with CR-pressure dominated atmospheres ($\Xcrinit \ge 3$). This is a result of the stabilising effect of the increasing CR pressure which smooths the CR pressure distribution and reduces the CR gradients. This suppresses the generation of Alf\'en waves and consequently reduces the CR scattering rate. According to Eq.~\eqref{eq:intrinsiccrdiffusioncoefficient}, CRs with such high diffusion coefficients scatter, on average, less than once per gigayear. Since the duration of our simulations is of order this timescale ($t_\mathrm{sim} \sim 1$ Gyr), these CRs are currently not scattered by Alfv\'en waves and the CR force parallel to the magnetic field line is suppressed. We note that this is a temporary state because these CRs can migrate to regions with lower diffusion coefficients, where they resume to scatter with Alfv\'en waves and exert the parallel forces again.

Second, in regions of efficient CR scattering, the distribution of $\kcr$ shifts toward smaller values with increasing $\Xcr$. This behaviour is in line with the self-confinement picture of CR transport: in regions with higher $\Xcr$, the greater abundance of CRs amplifies the generation of gyroresonant Alfv\'en waves. These waves, in turn, increase CR scattering rates, confining CRs more effectively and thereby reducing their effective diffusivity. This interplay between CR abundance, wave generation, and scattering underscores the critical role of self-confinement in regulating CR transport properties across the CGM.

As stated above, $\kcr$ only captures the diffusive motion of CRs while ignoring their streaming motion. Therefore, it is not comparable to diffusion coefficients that arise from diffusion-only models. To mitigate this, we introduced the effective CR diffusion coefficient \citep{Thomas2023},
\begin{equation}
    \keff = \frac{\fcr}{\bs{b} \bcdot \bnabla \ecr},
    \label{eq:kappa_eff}
\end{equation}
which characterises the entire CR transport, streaming or diffusing, using a single quantity. Here, the term $\bs{b} \bcdot \bnabla \ecr$ is the gradient of CR energy density along the magnetic field and $\fcr$ is the CR energy flux that is evolved in our simulations as one of the CRMHD quantities. Defining such an effective CR diffusion coefficient is useful because in the limit of a purely diffusive CR transport, both the intrinsic and effective diffusion coefficient are the same (i.e.,\ $\kcr = \keff$). In general, the effective quantity $\keff$ is the diffusion coefficient needed to mimic the actual CR transport in a hypothetical pure-diffusion model. Figure~\ref{fig:kappa_eff} illustrates the two-dimensional CR distribution in $\keff$ and $\Xcr$ at $t=6\,\tcool$. The distribution spans an immense range, from low values around $\keff \sim 10^{25}\,\unitKappa$ to highly diffusive CR populations with $\keff \sim 10^{35}\,\unitKappa$, covering ten orders of magnitude. There is a noticeable trend to higher values of $\keff$ for increasing $\Xcr$.

\begin{figure*}
    \centering
    \includegraphics[width=8.5cm]{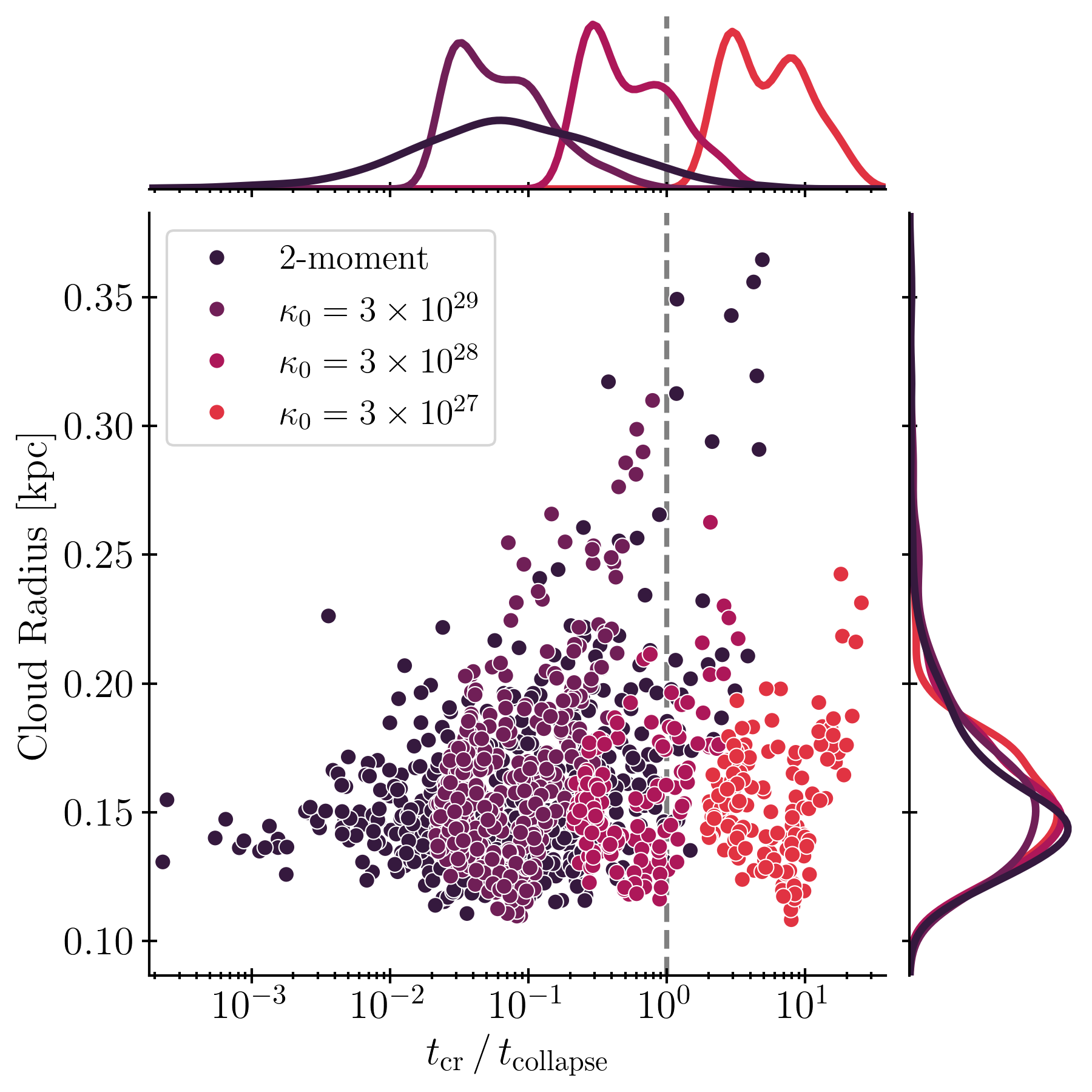}
    \includegraphics[width=8.5cm]{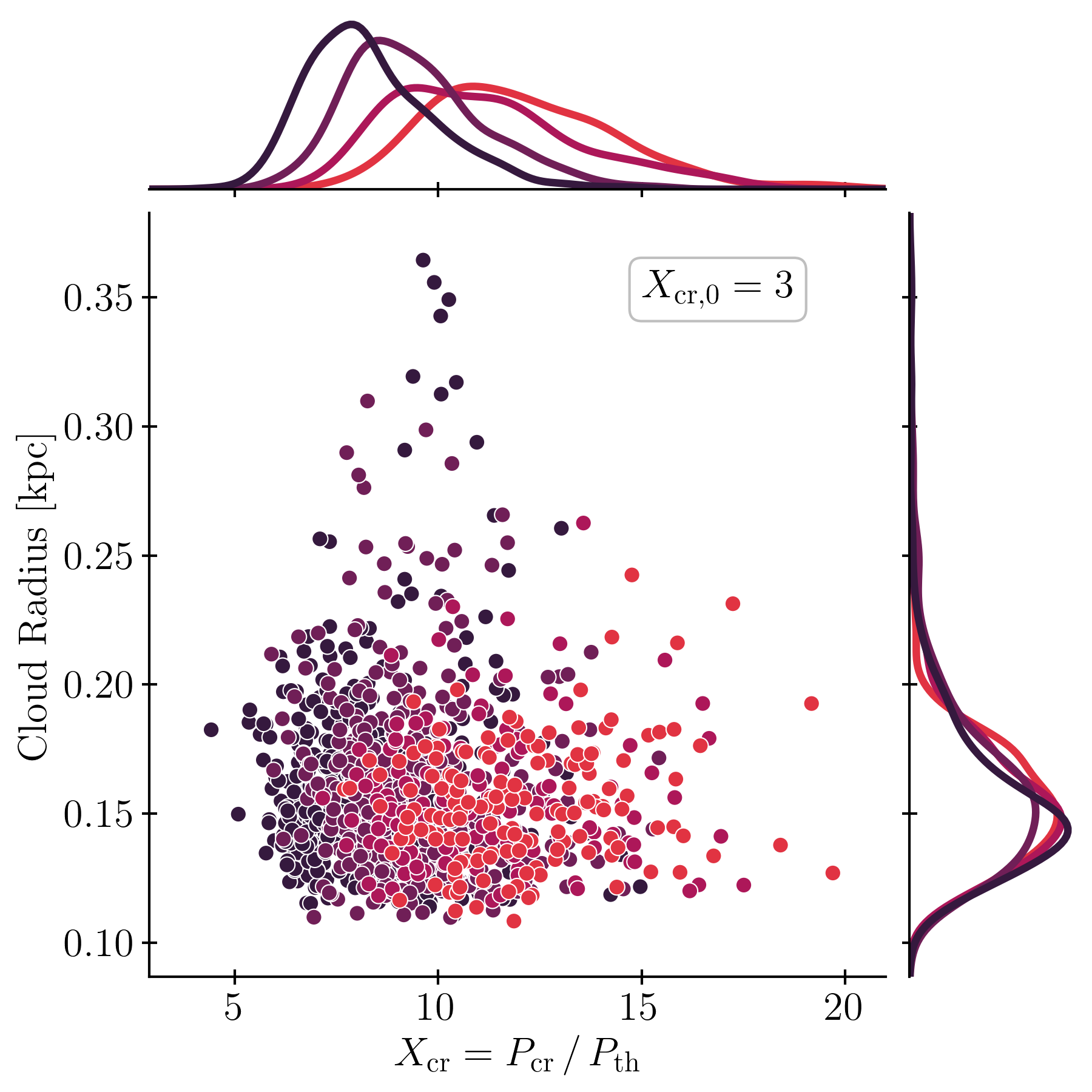}
    \caption{Ratios of the CR transport timescale to the cloud collapse timescale (left) and the relative CR pressure within cold clouds (right), depicted for various cloud radii. The 1D probability density of the corresponding variable is displayed above each axis with a linear scaling. We analyse cold gas clouds within the region $-0.25H \leq x, y \leq 0.25H$ and $0.25H \leq z \leq 1.75H$\protect\footnotemark, considering eight consecutive snapshots following the identification of the first cloud by the cloud finder. Different colours represent different CR transport models (as detailed in the figure legend). All simulations are initialised with $\Xcrinit=3$. The dashed vertical line in the left panel separates regions where clouds collapse is slower ($t_\mathrm{cr} /  t_\mathrm{collaspe} < 1$) or faster ($t_\mathrm{cr} /  t_\mathrm{collaspe} > 1$) than CRs escape these clouds. Fast transport enables CRs to escape collapsing clouds on timescales shorter than the cloud collapse time, reducing their relative pressure. In contrast, slow transport traps CRs within collapsing regions, increasing their pressure support.}%
    \label{fig:tcr_tcool}
\end{figure*}

To represent this finding with a single metric, we examine the CR-energy-weighted effective diffusion coefficient, $\Bar{\kappa}_\mathrm{eff}$, which emphasises the dominant contribution of high-energy CRs. At each snapshot between $1\,\tcool$ and $7\,\tcool$, we determine the CR-energy-weighted median of $\keff$ and we compute the mean across these values. Even for simulations with moderate CR pressure support (i.e.,\ $\Xcrinit=0.03$ and $\Xcrinit=0.3$), the value of $\Bar{\kappa}_\mathrm{eff}$ is increased by a factor of $\sim$10 in comparison to the ``canonical'' ISM CR diffusion coefficient in diffusion-only models. For the two cases, we find $\Bar{\kappa}_\mathrm{eff}=1.3 \times 10^{29}\,\unitKappa$ and $\Bar{\kappa}_\mathrm{eff}=1.6\times 10^{29}\,\unitKappa$, respectively. The CR-dominated runs with $\Xcrinit=3$ and $\Xcrinit=30$ result in even higher values of $\Bar{\kappa}_\mathrm{eff}=4.3 \times 10^{29}\,\unitKappa$ and $\Bar{\kappa}_\mathrm{eff}=8\times 10^{29}\,\unitKappa$, respectively. In the following, we utilise $\Bar{\kappa}_\mathrm{eff}$ to examine the impact of the effective CR transport speed on TI.

\subsection{Non-Fickian cosmic ray diffusion}
The simulations in the previous section employed the two-moment framework for CR transport and showed a highly variable effective CR diffusion coefficient and CR transport speed that depend on their environment in the TI-affected CGM. Quantifying the impact of the CR transport speed on the TI-induced cloud collapse becomes difficult due to the simultaneous presence of both transport modes. To simplify the analysis while maintaining the possibility for direct comparison, we conduct a series of simulations employing purely diffusive CR transport, with a constant CR diffusion coefficient, $\kappa_0$. A standard approach to describe such diffusion-only models is through Telegrapher's equation \citep{Gombosi1993, Litvinenko2013, Thomas2019}:
\begin{equation} 
\label{eq:telegraph} 
    \tau_\mathrm{cr} \frac{\partial^2 \ecr}{\partial t^2} + \frac{\partial \ecr}{\partial t} = \bnabla\bcdot\left(\kappa_0 \bs{b} \bs{b} \bcdot \bnabla\ecr \right), 
\end{equation} 
where $\kappa_0$ is the constant CR diffusion coefficient, and $\tau_\mathrm{cr}=3\kappa_0/c^{2}$ defines the CR diffusion timescale. The Telegrapher's equation extends the usual Fickian diffusion model by introducing a second-order time derivative, effectively accounting for the finite propagation speed of CRs. Unlike the purely diffusive approach, which assumes instantaneous diffusion, the Telegrapher's equation describes CR transport as a advection-like propagation combined with diffusion. This non-Fickian description of CR transport \citep{Malkov2015, Litvinenko2016, Rodrigues2019} results in a hyperbolic equation, for which a numerical discretisation in the context of finite volume schemes is more straightforward in comparison to the parabolic anisotropic diffusion equation. Furthermore, we adapted our solver for the two-moment equations to simulate this constant diffusion coefficient model. This allows for a straightforward comparison between simulations using the non-Fickian approximation with a constant diffusion coefficient and the two-moment self-confinement model. The underlying physical approximation of the two models is different, and there is no satisfactory explanation for the constant diffusion model in the self-confinement picture of CR transport. The absence of explicit coupling to Alfv\'en waves means that scattering must be assumed to originate from an external process, not driven by the CRs themselves. Thus, in the diffusion picture, the motion of CRs along magnetic field lines is in the direction opposite to the CR pressure gradient (see the right-hand side of Eq.~\ref{eq:telegraph}).

\begin{figure}
    \centering
    \resizebox{\hsize}{!}{\includegraphics{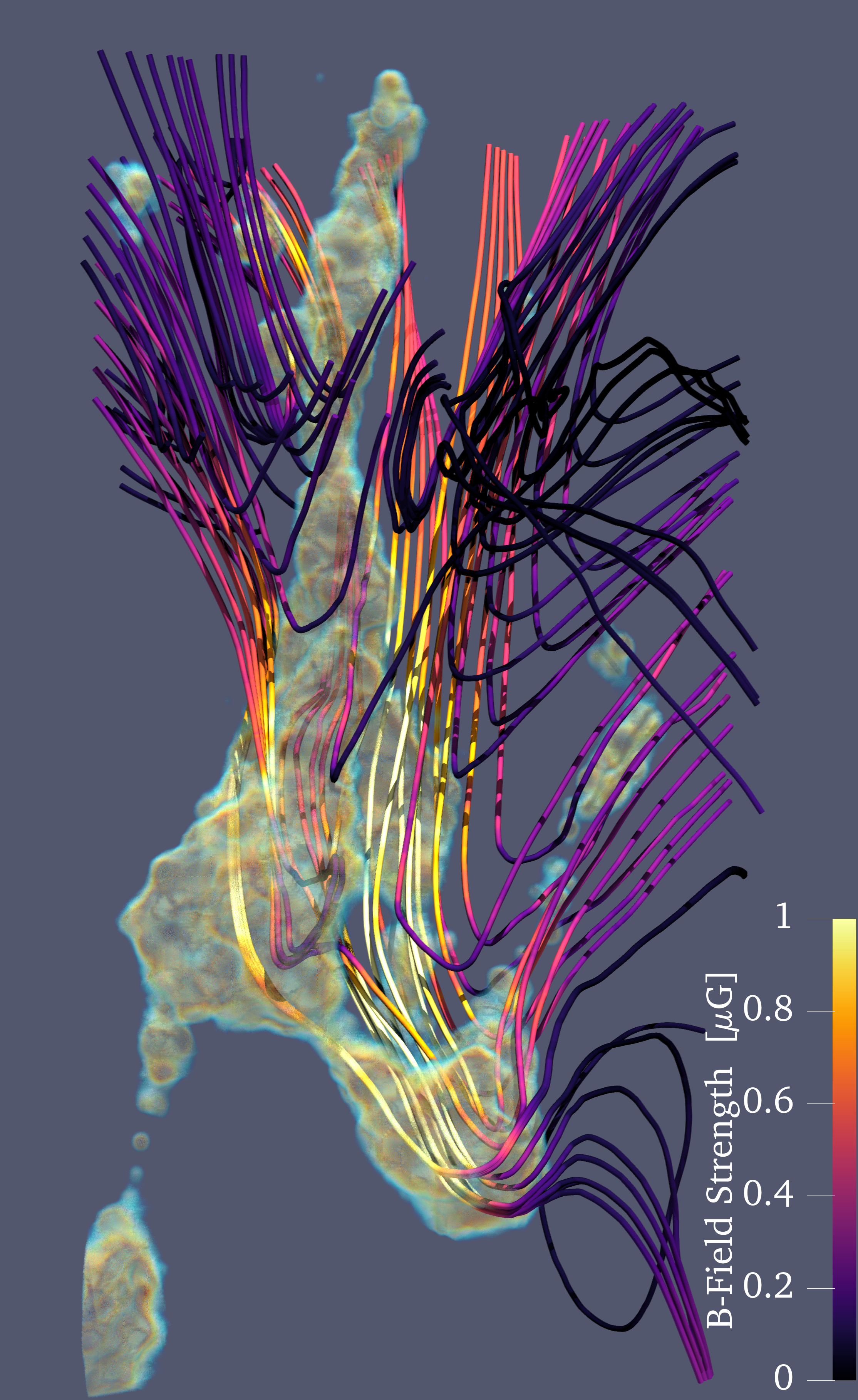}}
    \caption{Visual impression of the magnetic field topology in and around one of the clouds. This cloud has formed through TI and is currently falling down. The volume rendering highlights dense gas and the magnetic field lines are coloured according to the local magnetic field strength. The shown magnetic field lines connect the inside of the cloud to the ambient gas above the cloud. The magnetic field inside the cloud is stronger (achieving $B \sim 1\,\mu$G) compared to the magnetic field in the ambient gas.}%
    \label{fig:magneticfieldlines}
\end{figure}

To investigate the influence of CR transport speed on TI, we conducted a subset of simulations from Sec.~\ref{sec:2momentcrs}, this time utilising the diffusion-only model for the transport of CR energy. As before, all simulations are initialised with $\tau=0.3$, $\Xmaginit=0.01$, and $\Xkininit=0.3$, and each run is evolved for eight cooling times. Additionally, we use three different values for the constant CR diffusion coefficient each representing a different regime: slow diffusion with $\kappa_0 = 3\times10^{27}\,\unitKappa$, the ``canonical'' value ($ \kappa_0 = 3\times10^{28}\,\unitKappa$), and fast diffusion with $\kappa_0 = 3\times10^{29}\,\unitKappa$, and compare the results to a run employing two-moment CR transport.

Figure~\ref{fig:crtransports} demonstrates that fast and active CR transport accelerates the onset of collapse. The figure shows projections of the hydrogen column density, $\NH$. All simulations are initialised with a relative CR pressure of $\Xcrinit=3$, but different CR transport mechanisms. The snapshots are taken at $t=6\,\tcool$. There is a consistent trend of increased gas condensation with higher (effective) CR diffusion coefficient. In the purely advective case, no cold gas forms, as discussed in Sec.~\ref{sec:advectivecrs}. However, cold gas starts to precipitate already for slowly diffusing CRs. This shows that as soon as there is an escape channel for CRs out of the collapsing TI perturbations, these perturbations lose the CR-pressure support that is present in the overly confining CR-advection models. 

As the (effective) CR diffusion coefficient increases, the simulations show a gradual increase in dense gas formation. This condensation is fastest in the two-moment transport model. Notably, the diffusion model with $\kappa_0 = 3\times10^{29}\,\unitKappa$ produces results comparable to those of the two-moment transport setup. We attribute this similarity to the close match of the constant diffusion coefficient in this simulation to the effective diffusion coefficient of the two-moment model, $\keff = 4.3\times10^{29}\,\unitKappa$. Consequently, both models exhibit analogous CR transport behaviour and cold gas condensation patterns.

Figure~\ref{fig:crtransport_coldmass} confirms these findings in a more quantitative way. In each of the four metrics we observe a clear transition from the two-moment model, which represents the scenario of fastest CR transport, to the advective model, constituting fully ineffective CR transport. The two-moment model exhibits a nearly identical evolution to the fast-diffusion model across all metrics. By $t=6\,\tcool$, both models have generated significantly more cold gas compared to the setups with slower diffusion. Specifically, they produce approximately ten times more cold gas than the model with $\kappa_0 = 3\times10^{27}\,\unitKappa$ and five times more than the model with $\kappa_0 = 3\times10^{28}\,\unitKappa$. This further supports our earlier finding that higher CR transport velocities play a critical role in facilitating cold gas condensation.

\footnotetext{We focus on this reduced region to avoid double counting clouds across periodic boundaries and to ensure consistent analysis across resolutions (see Sec.~\ref{sec:resolution}) because the memory-intensive cloud finder makes analysing the full domain computationally very expensive at high resolutions.}

\subsection{Escape of cosmic rays from collapsing clouds}
Cosmic rays can counteract TI by providing additional pressure support but only as long as they remain trapped within the collapsing regions so that they can build up a pressure gradient. Therefore, when the confinement time of CRs inside cold clouds is prolonged, their influence on TI is expected to be more pronounced. To quantify this effect, we defined the average time required for CRs to escape from a cold cloud as 
\begin{equation}
    \label{eq:crdiffusiontimescale}
    \tcr = \frac{r^2_\mathrm{cloud}}{\kappa},
\end{equation}
where $r_\mathrm{cloud}$ is the effective cloud radius (see Sec.~\ref{subsec:coldcloudidentification} for the definition), and $\kappa$ is either the effective CR diffusion coefficient in two-moment CRMHD ($\keff$) or the pure diffusion coefficient in our non-Fickian diffusion model ($\kappa_0$). For this definition, we continue to approximate the clouds as spheres and assume that CRs diffuse freely through the clouds with their respective diffusion coefficients.

In the absence of self-gravity, the collapse timescale of a cold cloud is typically governed by the cooling timescale, $\tcol \sim \tcool$, as the cloud's internal dynamics is primarily driven by radiative cooling and pressure balance rather than gravitational collapse. In this context, $\tcool$ sets the pace at which the cold cloud loses energy and contracts due to thermal pressure imbalance. The cloud will collapse as long as it cools faster than it can regain pressure support. For typical conditions in the CGM, where cold clouds are embedded in a hot, diffuse medium, $\tcool$ ranges from several kyr to hundreds of Myr, depending on the cloud's density, temperature, and metallicity. 

To compute the cooling time of a cold cloud, we adopt its mean density, 
$ \rho_\mathrm{cloud} = M_\mathrm{cloud}\,/\,V_\mathrm{cloud}$,
and mean internal specific energy,  $U_\mathrm{th, cloud} = \sum_i (m_i\,u_i)\,/\,M_\mathrm{cloud}$,
as representative values. Here, $M_\mathrm{cloud}$ and $V_\mathrm{cloud}$ are the total mass and the total volume of the cloud, respectively, while $m_i$ and $u_i$ denote the mass and the internal specific energy of the computational cell $i$ within the cloud. By averaging the cloud's thermodynamic properties, we account for both the faster cooling of the dense, cold core and the slower evolution of the outer layers, characterised by lower densities and higher temperatures in an average sense.

When $\tcr < \tcol$, CRs escape from the cold clouds rapidly enough that their impact on the collapse process is minimal. Conversely, when CR transport occurs over timescales longer than the collapse time ($\tcr \gtrsim \tcol$), CRs can set up a pressure gradient that points toward the cloud centre, which is able to significantly counteract TI to eventually halt cloud collapse.

\begin{figure*}
    \centering   
    \includegraphics[width=17cm]{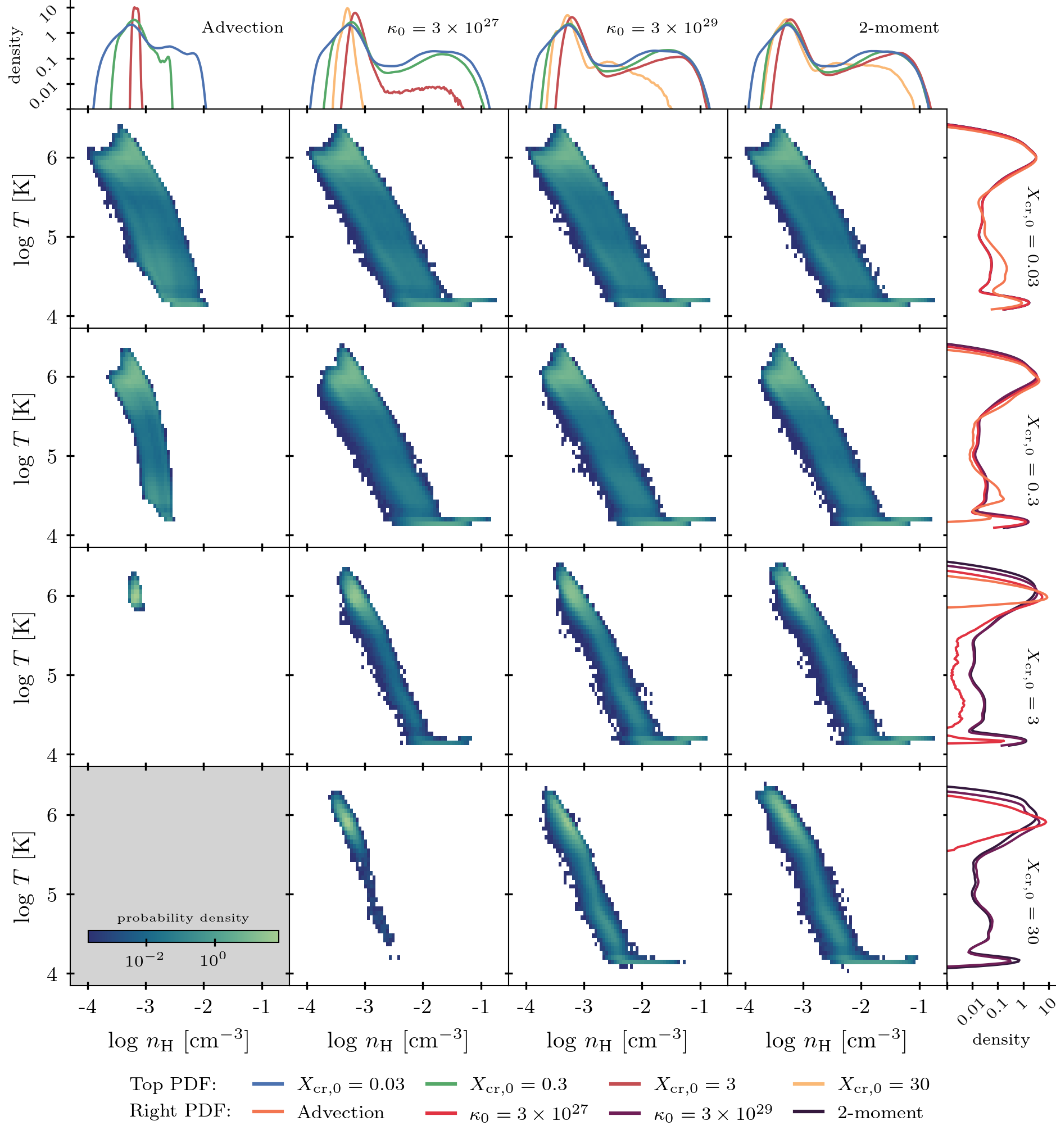}    
    \caption{Mass-weighted $\nH$-$T$ diagrams from our simulation suite. Columns represent different CR transport methods, while rows correspond to varying initial CR pressure fractions, $\Xcrinit$. The 1D probability densities of the respective variables are shown above each axis: columns reflect the resulting $\nH$ distributions for a fixed CR transport method across different $\Xcrinit$ and rows show the resulting $T$ distributions for a fixed $\Xcrinit$ across different CR transport methods. The phase diagrams are constructed using 70 bins of equal (logarithmic) width for both variables. All simulations are analysed at $t = 6\,\tcool$.}%
    \label{fig:phasediagram}
\end{figure*}

\begin{figure*}
    \centering
    \includegraphics[width=17cm]{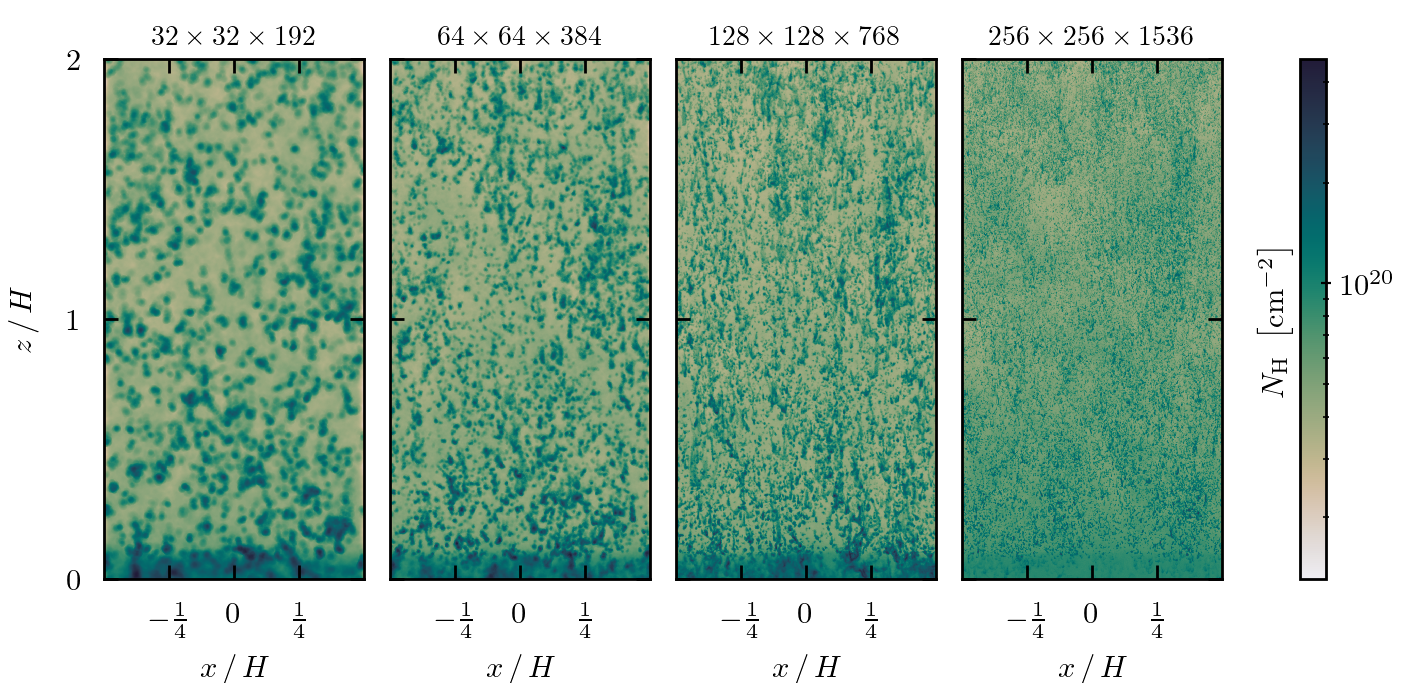}
    \caption{Projection showing the hydrogen column density, $\NH$, after the simulations run for $8\,\tcool$. From left to right, we plot the results from simulations with increasing resolution where each simulation employs two-moment CR transport and is initialised with $\Xcrinit=3$. The sizes and number of the cold clouds significantly depend on resolution.}
    \label{fig:proj_resolution}
\end{figure*}

Figure~\ref{fig:tcr_tcool} illustrates the impact of varying CR transport velocities on CR-cloud interactions. The left panel shows the distribution of $\tcr / \tcol$, while the right panel presents $\Xcr$, both evaluated for various cloud radii. The analysis includes eight consecutive snapshots starting from the moment the cloud finder identifies the first cloud. Two consecutive snapshots are separated by $\Delta t = 10$ Myr, which is significantly larger than the typical values of $\tcool \sim 0.5$ Myr and $\tcr \sim 0.02 \, \left(3 \times 10^{29}\,\unitKappa\,/\,\keff \right)$~Myr. This indicates that the clouds in the analysed snapshots are statistically independent. Consequently, some clouds may appear multiple times in the dataset, each data point sampling a different episode in their evolution with a different radius or thermodynamic state. The grey dashed line in the left panel represents the threshold where the CR diffusion time equals the collapse time: clouds to the left of this line enable CRs to escape faster than the collapse time of the cloud, whereas clouds to the right can retain CRs, which then slow down (or even halt) cloud collapse.

The timescale ratio $\tcr / \tcol$ exhibits a broad range, spanning $\lesssim 10^{-3}$ to $\gtrsim 20$. This variability arises from three factors: first, differences in cloud collapse timescales, $\tcol$, governed by internal thermodynamic properties for individual clouds; second, variations in cloud radii affecting $\tcr \propto r_\mathrm{cloud}^2$; and third, in simulations with two-moment CR transport, a wide range of $\keff$ values influences $\tcr \propto \keff^{-1}$.  

Fast CR transport in the two-moment and diffusion model with $\kappa_0 = 3 \times 10^{29}\,\unitKappa$, enables rapid CR escape and yields a median $\tcr / \tcol \sim 6 \times 10^{-2}$. For lower $\kappa_0$, slower CR escape increases the median ratio that reaches $\tcr / \tcol\sim 0.8$ ($\sim$7) for $\kappa_0 = 3 \times 10^{28}\,\unitKappa$ ($\kappa_0 = 3 \times 10^{27}\,\unitKappa$).  This behaviour influences CR pressure support during collapse as shown in the right panel. Slow CR transport traps CRs and enhances the pressure difference between the cloud to its surroundings and delays collapse. In contrast, fast CR transport reduces the CR pressure gradient and lowers $\Xcr$ within clouds.

\subsection{Cosmic ray motorways}
While our previous analyses suggest that CRs can be transported rapidly enough to escape contracting clouds without exerting significant pressure to resist collapse, the exact nature of their escape pathways remains unclear. CR transport is inherently anisotropic as CRs primarily drift along magnetic field lines either by streaming or diffusion. The efficiency of CR escape therefore depends critically on the magnetic field topology in and around the collapsing region because open field lines facilitate rapid CR transport, whereas closed or convoluted lines impede it. 

In Fig.~\ref{fig:magneticfieldlines}, we visualise the magnetic field topology surrounding one of the cold clouds from our simulations (which represents a typical situation as we confirmed). The magnetic field strength is considerably amplified inside the cloud ($\sim 1\,\mathrm{\mu G}$) compared to the ambient gas ($\sim 0.1\,\mathrm{\mu G}$) because the magnetic field lines are flux-frozen into the collapsing gas: when a cloud contracts and its volume decreases then the field lines are compressed and the magnetic field strength increases in proportion to the decrease in cloud size. The magnetic field lines permeate the cloud, align with its morphology, and guide the CR motion. Within the collapsing cloud, these field lines remain open and provide a clear and direct pathway -- essentially a ``CR motorway'' -- from the cloud's interior to the ambient medium. This open field line structure facilitates the efficient escape of CRs and allows them to stream or diffuse outward along these magnetic field lines, bypassing the dense, collapsing gas. These direct paths enable CRs to rapidly escape, thus preventing significant pressure build-up inside the cloud, and reducing the CR-mediated influence on TI.

\subsection{Impact of cosmic rays on gas phases}
Figure~\ref{fig:phasediagram} presents the mass-weighted $\nH$-$T$ diagrams for the entire simulation suite. The columns represent different CR transport models and the rows correspond to varying initial values of $\Xcrinit$. Within each row, differences highlight the impact of CR transport mechanisms under identical initial conditions and reveal their influence on gas phases. In contrast, differences within a column show the effects of increasing CR pressure support alongside changes arising from different initial conditions. Above each column and row, we provide 1D probability density distributions for the respective variables. This enables a clear comparison of CR transport models at fixed $\Xcrinit$ or across different $\Xcrinit$ for the same transport model. The colour gradient represents the probability density within each histogram bin, with all snapshots analysed at $t=6\,\tcool$. The results reveal several key trends:

\paragraph{Cosmic ray transport mechanisms.} Provided the CR pressure inside a collapsed cloud starts to dominate, CRs in a purely advective CR transport model suppress gas condensation in comparison to active CR transport models. Active CR transport can enhance gas condensation by up to a factor of $\sim$20 relative to advective models. However, for marginal CR pressure support ($\Xcrinit = 0.03$), both transport approaches produce similar temperature distribution profiles.

\paragraph{Impact of cosmic ray propagation speed.} Faster CR transport promotes the formation of denser and colder gas. This effect is particularly pronounced in CR pressure-dominated atmospheres ($\Xcrinit > 1$) with active CR transport, whereas differences in thermally dominated CGMs are minimal. Simulations using the two-moment model or a diffusion coefficient of $\kappa_0 = 3 \times 10^{29} \, \unitKappa$ yield nearly identical results.

\begin{figure*}
    \centering
    \includegraphics[width=0.49\linewidth]{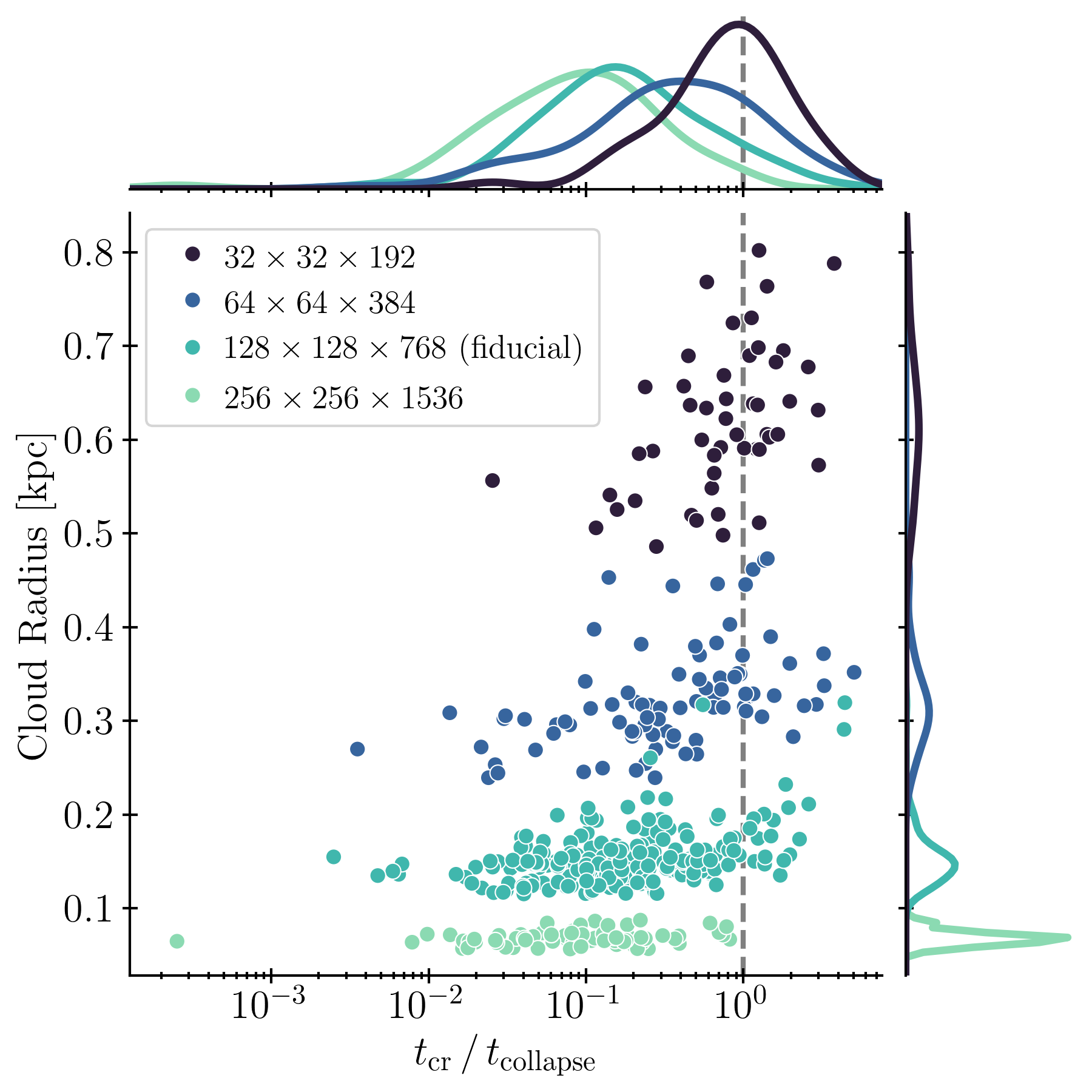}
    \includegraphics[width=0.49\linewidth]{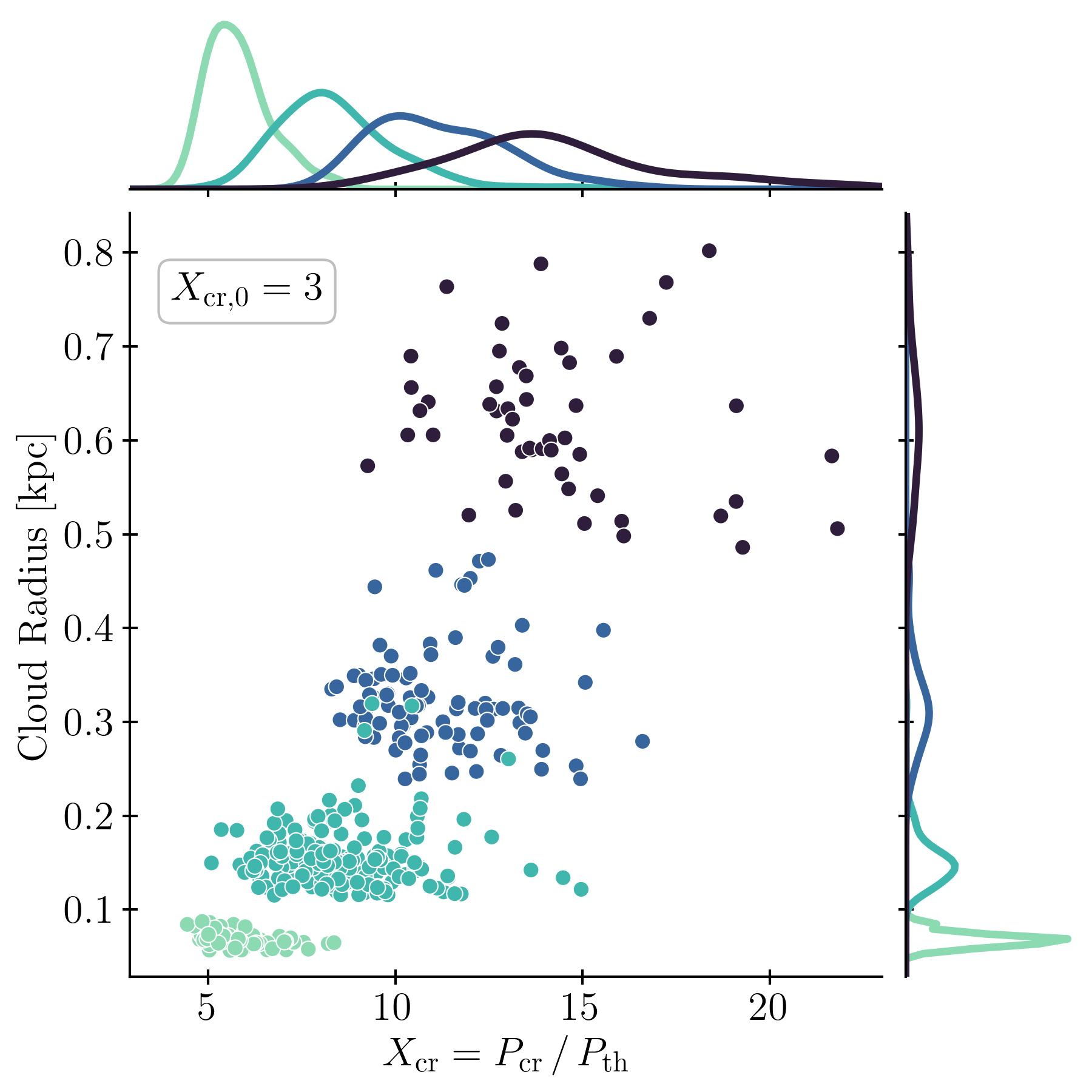}
    \caption{Same as Fig.~\ref{fig:tcr_tcool} but for different initial resolutions. All simulations employ two-moment CR transport and are initialised with $\Xcrinit=3$. A lower resolution produces larger clouds, inhibiting rapid CR escape and leading to enhanced CR pressure support within cold clouds.}
    \label{fig:tcr_tcool_resolution}
\end{figure*}

\section{Critical role of resolution}
\label{sec:resolution}
In this section, we examine the effects of numerical resolution in TI simulations of the CGM and focus on its implications for cloud sizes and the resulting effects on CR-cloud interactions.

\subsection{Theoretical cloud sizes}
Theoretical models predict that cold gas clouds should have characteristic sizes determined by the cooling length, $l_\mathrm{cloud} \sim c_\mathrm{s}\,\tcool$, where $c_\mathrm{s}$ is the local sound speed \citep{Voit1990, McCourt2018}. This relationship reflects the physical scale below which perturbations become thermally unstable and collapse into dense structures driven by cooling and compression by ambient pressure (instead of self-gravity) before significant thermal equilibration can occur. Under typical CGM conditions, this scale ranges from a few parsec to about a kiloparsec and depend on the gas density, temperature, and other environmental factors. 

Resolving this scale in global simulations of galaxies or halos is particularly challenging because it often lies well below the achievable spatial resolution. However, low resolution significantly suppresses turbulence-driven fragmentation, which is a crucial mechanism for forming dense, compact clumps from thermally unstable gas \citep{Scannapieco2015, Sparre2019}.
High-resolution simulations of isolated CGM patches have shown that resolving small-scale interactions between cold and hot phases is critical for accurately capturing cloud survival and dynamics \citep{Cooper2009, McCourt2015, Gronke2018, Sparre2020}. Without sufficient spatial resolution, simulations produce cold gas clouds that are overly smooth, diffuse, and unrealistically large when compared to observations. 
\citet{Fielding2020b} simulated radiative cooling layers and demonstrated that lower resolution simulations can exhibit significant pressure dips at intermediate entropy levels where cooling rates are highest. These pressure dips vanish with increased resolution and are caused by numerical diffusion, which artificially broadens the cooling layer. This artificial mixing of cool and hot phases creates intermediate-temperature gas instead of distinct cold clouds \citep{Hummels2019}. This impacts predictions for the column densities of low-ionisation species that are essential for tracing cold CGM gas in observations \citep{Peeples2019, Vandevoort2019}.

\subsection{Effect of resolution on cosmic ray-cloud interactions}
Resolution is equally crucial for accurately modelling CR transport because CR pressure gradients and interactions with gas and magnetic fields depend sensitively on small-scale structures. Low-resolution simulations often smooth out these gradients, which distorts the effective CR transport velocities and the forces CR exerted on the thermal gas. Resolving these gradients is essential for understanding how CRs influence the formation, evolution, and morphology of cold gas clouds in the CGM.

As discussed in Sec.~\ref{sec:crtransportspeed}, the size of simulated clouds significantly influences the role of CRs in shaping internal cloud dynamics. The timescale for CR escape from a cloud depends quadratically on its radius, $\tcr \propto r^2_\mathrm{cloud}$ (see Eq.~\ref{eq:crdiffusiontimescale}). Larger clouds confine CRs for extended periods, allowing their pressure forces to act on the cloud over longer timescales. This prolonged interaction amplifies the impact of CRs on the thermodynamic evolution of the cloud, potentially enhancing heating by CRs, altering cooling rates, and modifying condensation. Consequently, variations in cloud sizes can directly affect how CRs influence the stability, structure, and overall dynamical behaviour of cold CGM gas.

Figure~\ref{fig:proj_resolution} illustrates the impact of resolution on resulting cloud sizes. We show projections of the hydrogen column density, $\NH$, for simulations employing two-moment CR transport that are initialised with $\Xcrinit=3$. From left to right, the initial resolution of the simulations continuously increases. Higher resolution simulations show a higher number of more compact clouds or cloud complexes with visibly smaller associated cloud radii. 

A possible mechanism for this trend is that CRs can escape more rapidly from small cloud structures in the high-resolution simulations while large clouds, present in the low-resolution simulations, confine CRs for longer times. We verify this expectation in Fig.~\ref{fig:tcr_tcool_resolution}, which displays the ratio of CR escape time to cloud collapse time, $\tcr\,/\,\tcol$ (left panel), and the relative CR pressure, $\Xcr$ (right panel) for clouds of varying radii. Each cloud is colour-coded based on the initial resolution of the corresponding simulation. We show 1D probability densities for the respective variables above each axis. 
The highest-resolution run produces the smallest clouds, which facilitates rapid CR escape. In these simulations, the majority of clouds exhibit escape times shorter than their collapse times. Conversely, at lower resolutions, clouds are larger, which slows CR escape because the escape times become longer than the collapse times. This trend is particularly evident in the right panel where we observe that the CR pressure support increases for larger clouds formed in lower-resolution simulations, which amounts to $\Xcr$ values ranging from approximately ten to 22. By contrast, the high-resolution runs produce smaller, more compact clumps, resulting in lower CR pressure support and values of $\Xcr$ ranging from approximately four to eight. This comparison emphasises the significant influence of resolution on the interplay between CR dynamics and cloud formation in the CGM.

\section{Discussion and limitations}
\label{sec:discussion}

\subsection{Comparison to previous work}
In the pioneering study by \citet{Butsky2020}, the first comprehensive 3D MHD simulations investigating the role of CRs in the CGM were conducted using the \textsc{Enzo} code \citep{Bryan2014}. Their work explores the influence of CRs on the formation and evolution of cold gas driven by TI within the CGM. The simulation setup consists of a stratified box in hydrostatic equilibrium with an isocooling density profile. The initial magnetic field is oriented uniformly along the $x$-axis and has a magnetic-to-thermal pressure ratio of $\Xmag=0.01$. To initiate gas condensation, density perturbations are manually introduced in the initial conditions. Our setup is similar to that described by \citet{Butsky2020} but with some revisions. We incorporate turbulent velocity and magnetic fields in the initial conditions, such that fluctuations in these two fields do not need to be seeded from the dynamics started by density fluctuations. These initial turbulent fields naturally introduce density perturbations, negating the need for manually imposed density variations and better capturing the interplay between turbulence, magnetic fields, and CR dynamics in shaping cold gas structures. The turbulent magnetic fields also point in random directions starting from the beginning of the simulations, which leads to anisotropic CR transport that is not artificially correlated with one of the coordinate axes of the simulation box or the symmetry axis of the density and pressure profiles.

\citet{Butsky2020} find that CRs provide non-thermal pressure support, preventing cold gas from compressing as it cools, which results in lower densities in comparison to the pure MHD case. This effect allows cold gas to cool almost isochorically when CR pressure is dominant. Additionally, the presence of CR pressure allows cold gas clouds to reach larger sizes. In cases where CR pressure is relatively low, the cold gas forms as a `mist' of small, pressure-supported cloudlets, while cloud sizes can increase by orders of magnitude in CR-pressure-dominated halos with inefficient CR transport compared to purely thermally driven models. Our work supports these findings for the limiting case of purely advective CR transport. By incorporating the streaming and diffusion CR transport mechanisms, \citet{Butsky2020} demonstrate that CRs effectively redistribute pressure from dense cold clouds to the surrounding hot medium, leading to reduced cloud sizes and increased cold gas density and mass flux relative to adiabatic CR models. However, they also conclude that actively transported CRs have a substantial influence on cloud dynamics -- a finding that is not supported by our results, which suggests that cloud formation and evolution is governed by TI while CRs have at best a minor impact.

In Sec.~\ref{sec:resolution}, we discussed that insufficient resolution significantly affects the cloud-CRs interaction. This is because cloud radii are naturally limited by the numerical resolution, leading to larger CR escape times, which allows them to counteract TI-induced collapse for longer times. In their setup, \citet{Butsky2020} employ a static grid with a spatial resolution of 0.685 kpc per computational cell. Estimating that the smallest cold clouds span roughly three cells in diameter, the minimum cloud radius in their simulations is approximately $r_\mathrm{cloud} \sim 1 \, \mathrm{kpc}$. With their maximum CR diffusion coefficient of $\kcr = 7.9 \times 10^{29} \, \mathrm{cm^2 \, s^{-1}}$, the minimum CR escape time is $\tcr \sim 4 \times 10^{-4} \, \mathrm{Gyr}$. Using their cooling function and the temperature probability density function for the cold, dense phase at $T \sim 10^{4.75} \, \unitT$, we estimate the cooling timescale to be $\tcool \sim 2 \times 10^{-4} \, \mathrm{Gyr}$. Based on these parameters, we find that the ratio of the CR escape time to the collapse time is $\tcr / \tcool \sim 2$, indicating that CRs cannot escape collapsing regions quickly enough. This delayed escape maximises the non-thermal pressure support of CRs on the clouds. Even in their highest-resolution simulations, this ratio can be reduced by a maximum factor of two and is thus still above the parameter regime where CRs could escape faster than clouds are collapsing due to TI. From these considerations, we conclude that the limited cloud-growth observed by \citet{Butsky2020} can be explained by a CR-induced delay of cloud collapse, which is artificially promoted by insufficient resolution.

\citet{Tsung2023} investigate the impact of CR heating on TI in a CGM-like gravitationally stratified medium with vertical magnetic fields and two-moment CR transport utilising the approach of \citet{Jiang2018} in various 2D simulations with the \textsc{Athena++} code \citep{Stone2020}. In the linear phase, they confirm that CR heating causes entropy modes to propagate at a velocity proportional to the Alfv\'en speed, as predicted by theory \citep{Kempski2020}. However, this propagation, which could theoretically suppress TI, is limited to a narrow range of conditions where the ratio of gas cooling timescale and CR heating timescale approaches unity. The study’s main findings lie in the non-linear phase where TI leads to significant mass condensation and a cold disc formation in the mid-plane, altering CGM properties and phase structure.

\subsection{Limitations}
Our simulations are intentionally highly idealised, which is advantageous for isolating the specific influence of CRs on TI and their interactions with cold clouds. However, this approach lacks a fully realistic physical context that could affect cold gas formation and survival in the CGM. The heating model we use is a simplified representation of an isolated, dynamically stable, long-lived CGM, where we balance total cooling and heating globally on average. In a more physically realistic setup, local processes such as gas accretion, star formation, stellar and AGN feedback, and galactic mergers would naturally contribute to this equilibrium. This is particularly relevant in simulations evolved over several hundreds of Myr. On these timescales, the assumption of an isolated CGM becomes increasingly invalid as a consequence of the above processes \citep{Tumlinson2017, Hummels2019, Nelson2020}.

The initial density and temperature profiles in our simulations follow an isocooling relation ensuring a uniform initial cooling time across the computational domain. While this approach results in profiles that closely resemble isothermal ones, we do not intend to replicate realistic halo configurations with this setup. Instead, the primary motivation for using these profiles is to maintain consistency and comparability with previous studies that employed similar initial conditions and to allow for a more direct comparison of results. To maintain the global stability of the initial atmosphere while including additional CR pressure, the total mass in the system increases.

In our simulations, we adopt a fiducial ratio of $\tau = \tcool/\tff=0.3$, representative of regions in the CGM where rapid cooling promotes the condensation of cold gas. However, in reality, the CGM exhibits a wide range of thermodynamic conditions, with local cooling times varying significantly due to differences in density, thermal pressure, and metallicity. Even for fast CR transport models, there are likely to be dense, rapidly cooling regions of the CGM where $\tcr/\tcool \gtrsim 1$, leading to effective CR trapping. Thus, while our simulations focus on a specific cooling regime that favours the formation of multiphase gas, we expect a broad spectrum of CR–gas coupling efficiencies across the CGM, depending sensitively on the local cooling environment and corresponding collapse timescales.

The CR transport velocity in our simulation is governed by interactions between CRs and gyroresonant Alfv\'en waves alongside the processes that drive the growth and damping of these waves. Specifically, we consider the gyroresonant interactions of CRs and the non-linear Landau damping of Alfv\'en waves \citep{Kulsrud2005} because these are generally regarded as the primary influences on CR transport. However, additional damping mechanisms including ion-neutral damping \citep{Kulsrud1969}, linear Landau damping \citep{Wiener2018}, and turbulent damping \citep{Farmer2004, Lazarian2016, Holguin2019} may act in realistic systems to limit Alfv\'en wave amplification to reduce the CR scattering rate. This reduction, however, results in less confined CR populations with effectively higher CR transport velocities and positions the results of this paper as a lower bound on the CR escape timescale from collapsing regions and an upper bound on their pressure support in the cold phase. 

Our simulations are incapable of fully resolving the cooling length, $l_\mathrm{cool} = c_\mathrm{s}\,\tcool$ as demonstrated in Appendix~\ref{app:coolinglengthresolution}. While this limitation introduces some uncertainty in capturing the smallest scales of TI and cloud fragmentation \citep{McCourt2018}, we believe our key findings remain robust, particularly regarding the CR escape timescale from cooling clouds. Resolving $l_\mathrm{cool}$ is important for accurately capturing small-scale structures in the cold gas. However, our results consistently show that the dominant processes governing CR transport and their interaction with cooling clouds are largely determined by larger-scale dynamics and the interplay between the CR escape timescale and the cloud collapse timescale. This suggests that our conclusions regarding CR escape and its effects on cold gas dynamics are not strongly dependent on resolving $l_\mathrm{cool}$.

In addition, our simulations do not fully resolve the small-scale structure of the magnetic field within cold clouds and their turbulent boundary layers. Capturing these scales could impact CR transport. 
The turbulent boundary layers between the cold clouds and the surrounding hot, diffuse medium generate complex density and pressure gradients, which enhance CR scattering and thereby reduce their effective transport speed \citep{Yan2004}. These boundary layers can therefore act as bottlenecks for CRs \citep{Wiener2017b}, limiting their escape from the cloud interior and causing them to accumulate near the cloud edges. This accumulation increases the local CR pressure, which in turn exerts an outward force on the cloud outskirts. Moreover, magnetic fields within these turbulent layers are amplified and become more organised within cold gas, often forming filamentary structures \citep{Zhao2023, Das2024}. While such local coherence may promote CR streaming along field lines, the overall magnetic topology across the multiphase medium remains complex. The suppression of turbulent mixing and increased magnetic pressure reduce the connectivity between different gas phases, potentially confining CRs within cold regions and impeding their large-scale transport.

\section{Conclusions}
\label{sec:conclusions}
In this paper, we have explored the impact of CRs on the formation and evolution of cold clouds in the CGM. Utilising the moving-mesh code \textsc{Arepo}, we performed 3D CRMHD simulations of idealised CGM environments. These simulations started from initial conditions characterised by an isocooling gas stratification, a turbulent velocity field, a turbulent magnetic field, a realistic gas cooling function, and a global heating function to replicate heating from galactic feedback. We incorporated CRs as a constant fraction of the thermal pressure, investigated various CR transport mechanisms, and assessed their significance. Our key findings are outlined as follows:

\begin{enumerate}
    \item Purely advective CRs significantly alter the morphology of the CGM. The additional non-thermal pressure support from CRs either inhibits the collapse of thermally unstable regions or delays their collapse in comparison to purely thermal scenarios (cf.\ Fig.~\ref{fig:cradv}). The strength of this effect is proportional to the initial CR pressure in the simulation. However, purely advective CRs represent an academic scenario because realistic CRs are transported down their pressure gradient along magnetic field lines via streaming and diffusion in realistic astrophysical environments.

    \item Including CR streaming and diffusion fundamentally alters the evolution of a collapsing cold cloud. As CRs are no longer fully coupled to the thermal gas movement, their dynamics are predominantly dictated by the CR pressure gradient along the magnetic field lines. When a cold cloud collapses, not only is its gas compressed, but the CRs that are tied to the magnetic field lines are also compressed. This increases the CR pressure within the collapsing regions and creates a pressure gradient between the cloud interior and the surrounding medium. The magnetic field lines that penetrate the cloud serve as pathways for the CRs to escape the collapsing regions on very short timescales (cf. Fig.~\ref{fig:magneticfieldlines}). This counteracts the build-up of CRs inside the cloud and impedes the stabilising effect of the CR pressure. Consequently, CRs cannot prevent the cloud's collapse, and the morphology of the CGM nearly resembles that of a purely thermal setup. This conclusion is mostly independent of the initial CR pressure (cf.\ Fig.~\ref{fig:2momCRs}).

    \item The effective CR diffusion coefficient, $\keff$, exhibits a wide dynamic range, spanning several orders of magnitude from $10^{25}$ to $10^{34}\,\unitKappa$ (see Fig.~\ref{fig:kappa_eff}), and it is generally dependent on $\Xcr$. Given the significant variation in $\Xcr$ expected in realistic CGMs, this broad range underscores the importance of employing the two-moment CR transport method. This approach is able to capture the anisotropic and dynamic nature of CR propagation in these diverse conditions.

    \item The CR escape timescale from collapsing regions is dictated by the effective CR diffusion coefficient, $\keff$ (cf.\ Fig.\ref{fig:crtransports}), rather than by the CR pressure. When $\keff$ is close to or smaller than its canonical value in the ISM of the Galaxy ($\lesssim 3\times10^{28}\,\unitKappa$), CRs can sustain their pressure support inside the thermally unstable regions for a longer period, thereby delaying the onset of collapse. For higher values of $\keff$ ($\gtrsim 3\times 10^{29}$), CRs escape quickly from the cold clouds (cf.\ Fig.~\ref{fig:tcr_tcool}).

    \item Resolution plays a pivotal role in accurately assessing the influence of CRs on cold gas in the CGM. The size of condensing clouds is strongly influenced by the resolution of computational cells, with low-resolution simulations yielding clouds up to ten times larger than those formed in high-resolution runs (cf.\ Fig.~\ref{fig:proj_resolution}). Since the CR escape timescale depends quadratically on cloud size, $\tcr = r^2_\mathrm{cloud}/\keff$, larger clouds retain CRs for significantly longer durations. This extended confinement amplifies the CR impact on cloud dynamics (cf.\ Fig.~\ref{fig:tcr_tcool_resolution}).
\end{enumerate}

These findings collectively demonstrate that the dynamical role of CRs in the CGM is highly sensitive to their transport properties and to the numerical resolution of the simulations. Accurate modelling of CR transport, particularly through two-moment methods, is therefore essential for capturing their true impact on cold gas evolution and the resulting CGM morphology.

\begin{acknowledgements} We thank the anonymous referee for helpful comments that improved the clarity of this manuscript. We acknowledge support by the European Research Council under ERC-AdG grant PICOGAL-101019746. This work was supported by the North-German Supercomputing Alliance (HLRN) under project bbp00070. The projections and slices presented in this work were generated using \texttt{Paicos} \citep{Berlok2024}. Additionally, all Python-based analysis scripts relied on \texttt{matplotlib} \citep{Matplotlib}, \texttt{seaborn} \citep{Seaborn} and \texttt{cmocean} \citep{cmocean} for visualisation including colour maps, and \texttt{numpy} \citep{Numpy} for numerical computations. 

\end{acknowledgements}

\bibliographystyle{aa_url}
\bibliography{refs}

\begin{appendix}
\onecolumn

\section{Resolution of cooling length}
\label{app:coolinglengthresolution}
\begin{figure}[h!]
    \centering
    \includegraphics{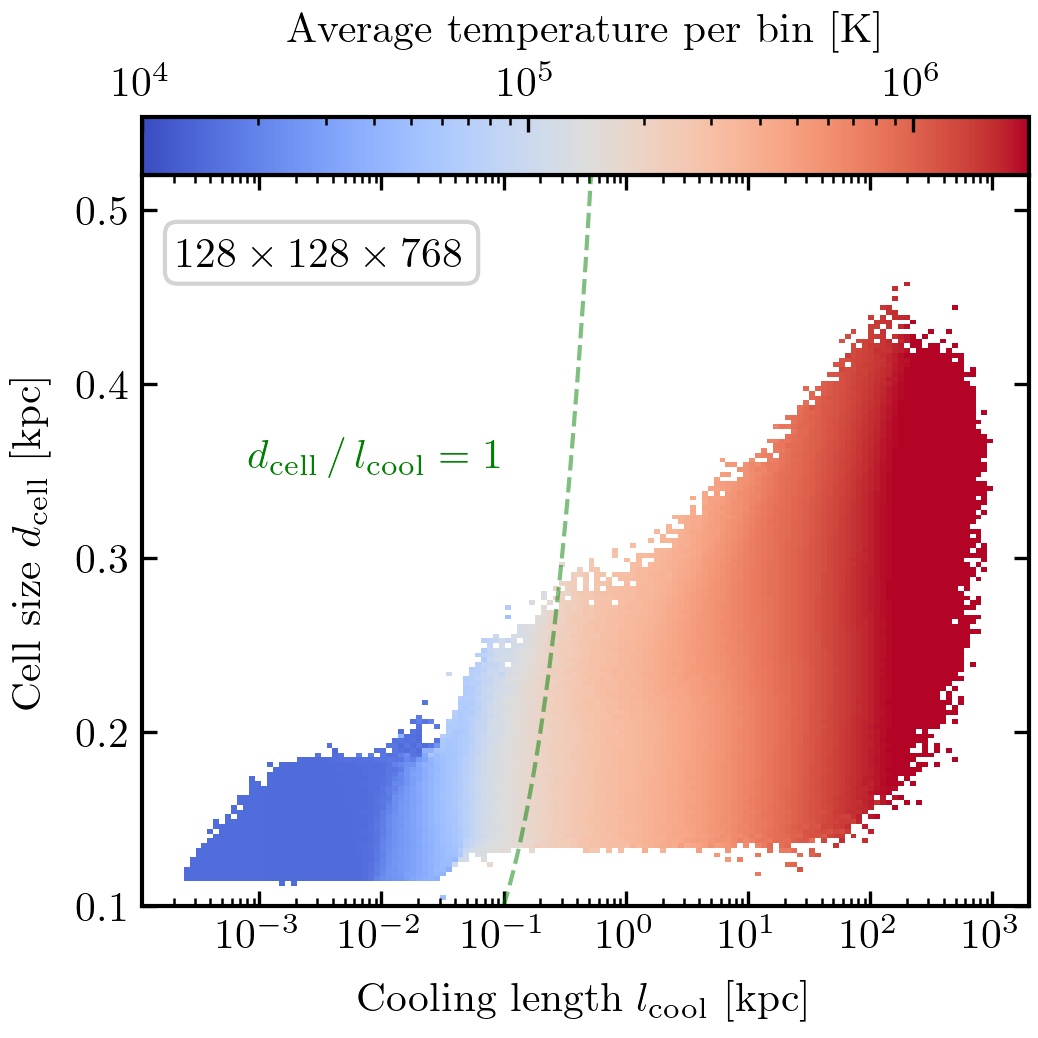}
    \caption{Two-dimensional histogram displaying the cooling length across various computational cell sizes of our fiducial simulation with $\tau = 0.3$ and $\Xcrinit = 3$. Each bin is colour-coded according to its average temperature. The green dashed line indicates where the cooling length equals the cell size, demonstrating that gas with $T \gtrsim 10^{5}\,\unitT$ is well-resolved in our simulations.}
    \label{fig:coolingLength}
\end{figure}

To highlight the impact of numerical resolution in our fiducial simulation with $\tau = 0.3$ and $\Xcrinit = 3$, we show the relationship between the cooling length, $l_\mathrm{cloud} = c_\mathrm{s}\,\tcool$, and the diameter of the computational cell, $d_\mathrm{cell} = (6 V_\mathrm{cell} / \pi)^{1/3}$, in 
Fig.~\ref{fig:coolingLength}. It shows a 2D histogram of the cooling length versus the cell size. Each bin is colour-coded by its average temperature, providing a detailed view of how resolution varies with the thermodynamic state of the gas. The green dashed line marks the thermodynamic states where the corresponding cooling length is equal to the cell size, delineating the boundary where thermal processes are adequately resolved. Our results show that gas with temperatures $T \gtrsim 10^{5}\,\unitT$ lies to the right of this line, indicating that our simulations properly resolve the cooling length for this temperature regime. Conversely, gas with $T \lesssim 10^{5}\,\unitT$ falls to the left of this line, suggesting that cooling processes in this regime may be under-resolved.

\section{Thermal instability without cosmic rays}
\label{app:tiwithourcrs}

\begin{figure*}[!ht]
    \centering
    \includegraphics[width=17cm]{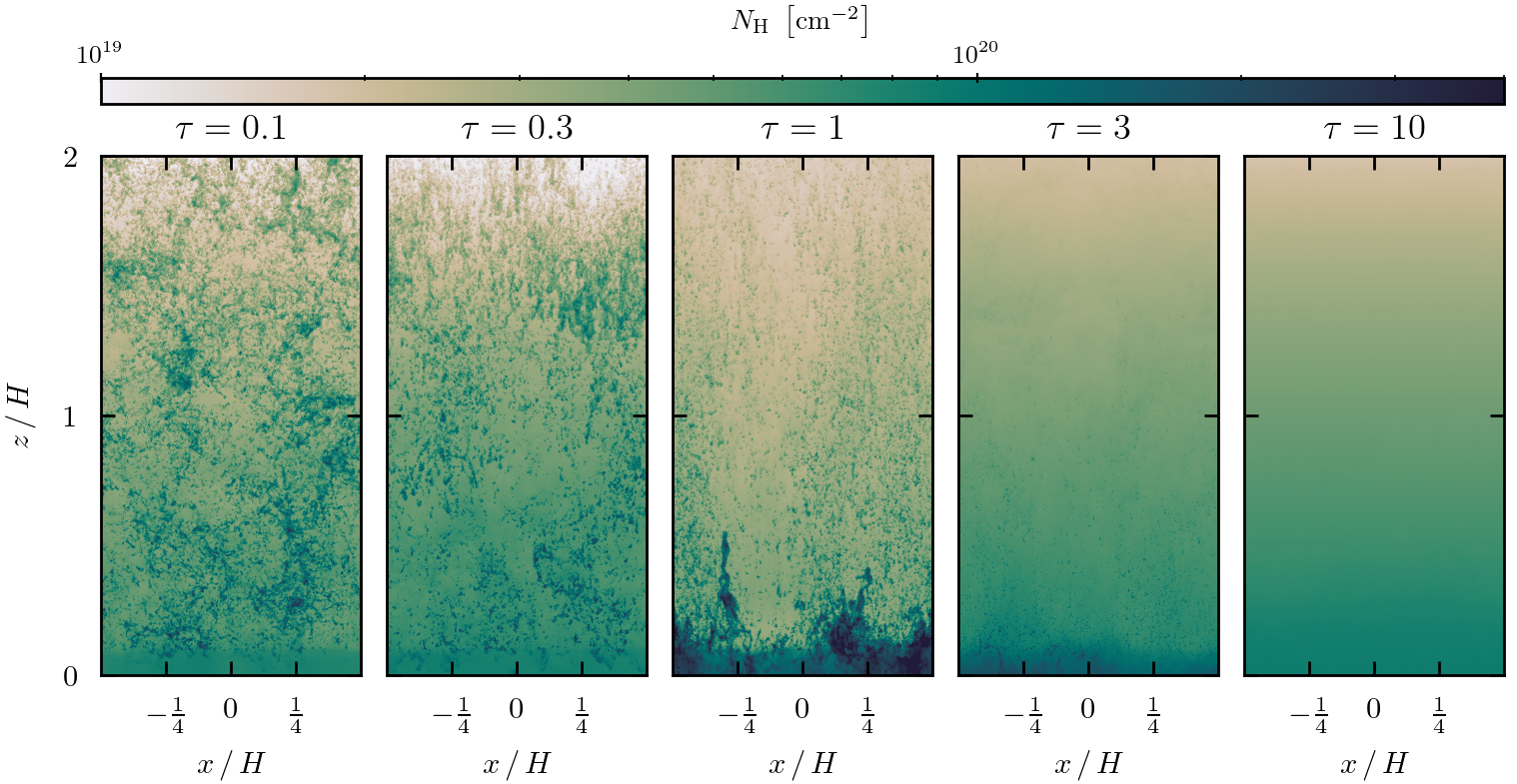}
    \caption{Projections showing the hydrogen column density, $\NH$, after the simulations run for $6\,\tcool$ without CRs. From left to right, we plot the results of the simulations employing different ratios of the cooling-to-free-fall time, $\tau$. From the figure, it is evident that this ratio plays a crucial role in determining the onset of TI.}%
    \label{fig:NoCRs}
\end{figure*}

\begin{figure*}[!ht]
    \centering
    \includegraphics[width=17cm]{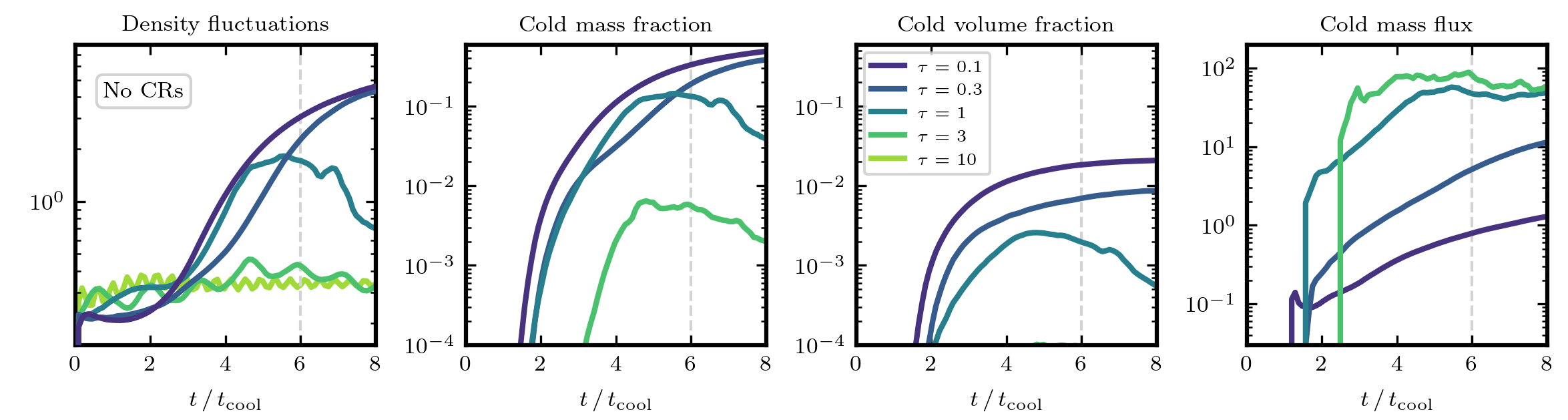}
    \caption{Time evolution of cold gas metrics for simulations without CRs. From left to right, we show density fluctuations, cold mass fraction, cold volume fraction, and cold mass flux. The different colours correspond to different values of $\tau$. We evaluate each computational cell in the region $0.25\,H \leq |z| \leq 2.75\,H$. The dashed vertical line denotes the time of the analyses shown in Fig.~\ref{fig:NoCRs}. }%
    \label{fig:coldGasMetrics_NoCRs}
\end{figure*}
In this appendix, we examine the onset of TI in the absence of CR physics. In their seminal work, \citet{McCourt2012} highlighted the critical importance of the cooling-to-free-fall time ratio, $\tau = \tcool / \tff$. They found that gas condensation occurs when $\tau \lesssim 1$. Since $\tcool$ is governed by the cooling function, changes in $\tau$ are achieved by modifying $\tff = \sqrt{2z / g(z)}$, where $z$ represents the vertical distance from the simulation box centre, and $g(z)$ denotes the gravitational acceleration, as defined in Eq.~\eqref{eq:gravity}. Specifically, we adjust the scale height, $H$, to increase or decrease $\tff$, thereby altering $\tau$.

Figure~\ref{fig:NoCRs} shows hydrogen column density projections, $\NH$, from simulations with identical initial density and temperature profiles but varying $\tau$. In simulations with smaller $\tau$, cold gas condenses rapidly from the background medium on timescales much shorter than $\tff$. As a result, a significant amount of cold gas remains suspended at high altitudes, even after multiple cooling cycles. Conversely, as $\tau$ increases, gravitational acceleration becomes the dominant process. Under these conditions, condensed gas clumps increasingly precipitate toward the gravitational centre, where they accumulate. In the case with $\tau=10$, we do not observe any TI. Notably, in contrast to \citet{Butsky2020}, we do not find an increase in cloud sizes with increasing $\tau$. We attribute this discrepancy to the significantly higher resolution of our simulations, which prevents artificial cloud coagulation caused by low-resolution-induced pressure gradients \citep{Gronke2020, Fielding2020b}. Overall, our findings are consistent with previous studies by \citet{McCourt2012}, \citet{Ji2018}, and \citet{Butsky2020}.

Figure~\ref{fig:coldGasMetrics_NoCRs} presents the cold gas metrics used in this study: density fluctuations, cold mass and volume fractions, and cold mass flux (from left to right). After the initial turbulence decays, density perturbations begin to grow in simulations with $\tau \lesssim 3$ at around $t \sim 2\,\tcool$. This leads to the condensation of cold gas from the hot ambient medium, increasing both the cold mass and the volume it occupies. For runs with $\tau \gtrsim 1$, these values decrease at later times due to cold gas precipitation out of the analysed domain. The simulation with $\tau=3$ produces significantly less cold gas because the shorter free-fall timescale allows cooling gas parcels to descend into higher-pressure regions near the gravitational centre earlier, leading to enhanced compressive heating of the cooling gas. Meanwhile, the cold mass flux increases with higher $\tau$ due to the faster velocities of cold gas driven by the shorter $\tff$.

\end{appendix}

\end{document}